\documentclass{article}
\usepackage{amsfonts}

\usepackage{graphicx}
\usepackage{amsmath}

\def\bea{\begin{eqnarray}}
\def\eea{\end{eqnarray}}

\begin{document}

%\vspace{1cm} \begin{flushright}
%CAMS/02-10
%\end{flushright}
%\vspace{1cm} \baselineskip=16pt

\begin{center}
\baselineskip=16pt \centerline{{\Large{\bf A Brief Introduction to
Particle Interactions \footnote{Series of lectures given at the
Summer School "Dirac Operators: Yesterday and Today" held at CAMS,
American University of Beirut, August 27-September 7, 2001.
Published by International Press, 2005.} }}} \vskip1 cm

Ali H. Chamseddine \vskip1cm
\centerline{\em Center for Advanced
Mathematical Sciences (CAMS) } \centerline{\em and}
\centerline{\em Physics Department, American University of Beirut,
Lebanon}
\end{center}

\vskip1 cm \centerline{\bf ABSTRACT}  I give a brief introduction to
particle interactions based on representations of Poincare Lie algebra. This
is later generalized to interactions based on representations of the
supersymmetry Lie algebra. Globally supersymmetric models with internal
symmetry and locally supersymmetric models leading to supergravity theories
are presented. I also discuss higher dimensional supergravity theories and
some of their applications. \vfill\eject
\bigskip
\tableofcontents
\section{Introduction}

Historically, the Dirac operator was put forward as a result of making the
Schr\"{o}dinger equation in quantum mechanics compatible with the special
theory of relativity. In modern physical theories, matter at very small
distances, or equivalenty, very high-energies, is made of elementary
particles whose interactions are the four fundamental forces in nature:
gravity, electromagnetic theory, weak-nuclear force and strong-nuclear force.

Locally the four-dimensional space-time is Minkowskian and physical states
are invariant under rotations (Lorentz transformations ) and translations.
The physical states  fall into representations of the Poincare algebra.
Particles will be characterized by their mass and spin. Matter particles
have spins $0$ and $\frac{1}{2}$ obeying respectively Klein-Gordon and Dirac
equations. Particles with spin-1 are the mediators of gauge interactions,
e.g. photons mediate the electromagnetic interactions between electrons
while all particles exchange gravitons for the gravitational interactions.

The general plan is to first study representations of the Poincare
algebra and place particles in these representations, and  to form
physical states obeying these symmetries and write
Lorentz-invariant interactions. The Poincare algebra is then
generalized to the supersymmetry algebra. We find representations
of the supersymmetry algebra and construct interactions invariant
under this new symmetry. For more details the reader may consult
the following references \cite{ACweinberg}, \cite{ACWest} ,
\cite{ACWess} , \cite{ACSohnius}, \cite{ACvan} , \cite{ACscherk} ,
\cite{ACduff} , \cite{ACwitten} .

\section{Poincare group}

Einstein's special theory of relativity states the equivalence of certain
inertial frames of reference. The coordinates $x^{\mu }$ in an inertial
frame satisfy
\begin{equation*}
ds^{2}=dx^{\mu }\eta _{\mu \nu }dx^{\nu }=dx^{\prime \mu }\eta _{\mu \nu
}dx^{\prime \nu },
\end{equation*}
where $x^{\mu }$ and $x^{\prime \mu }$ are the coordinates of the physical
states in different frames and $\eta _{\mu \nu }=diag\left(
1,-1,-1,-1\right) $ is the Minkowski metric. The transformations $x^{\mu
}\rightarrow x^{\prime \mu }=a^{\mu }+\Lambda _{\,\,\nu }^{\mu }x^{\nu }$
keeps the distance invariant provided that
\begin{equation}
\Lambda _{\,\,\kappa }^{\mu }\eta _{\mu \nu }\Lambda _{\,\,\lambda }^{\nu
}=\eta _{\kappa \lambda }.  \label{distance}
\end{equation}
Taking the determinant on both sides gives $\left( \det \Lambda \right)
^{2}=1$ or $\det \Lambda =\pm 1.$ In particular $\Lambda _{\,\,0}^{0}\eta
_{00}\Lambda _{\,\,0}^{0}+\Lambda _{\,\,0}^{i}\eta _{ij}\Lambda
_{\,\,0}^{j}=\eta _{00},$ or $\left( \Lambda _{\,\,0}^{0}\right)
^{2}=1+\left( \Lambda _{\,\,0}^{i}\right) ^{2}.$ This implies that $\Lambda
_{\,\,0}^{0}\geq 1,$ or $\Lambda _{\,\,0}^{0}\leq -1$ and transformations
are divided into four categories. The entire set of $\Lambda $ comprises the
homogeneous Lorentz group. The orthocronous Lorentz transformations are
charecterised by the condition $\Lambda _{\,\,0}^{0}\geq 1$ so that every
positive time-like vector transforms into positive time-like vector. If $%
\det \Lambda =1$ then this gives a group of restricted homogeneous
Lorentz-transformations. These are the only ones that could be obtained by a
continuous transformation of the identity. This is because it is not
possible to continously transform states with $\det \Lambda >0$ to states
with $\det \Lambda <0,$ or states with $\Lambda _{\,\,0}^{0}\geq 1$ to
states with $\Lambda _{\,\,0}^{0}\leq -1.$ Therefore these states correspond
to an $SO(3,1)$ group. The translations $a^{\mu }$ form a seperate group, an
abelian one, defined as $T\left( 4\right) .$ We write
\begin{align*}
x^{\prime }& =\Lambda x+a\equiv g\left( a,\Lambda \right) x, \\
x^{\prime \prime }& =\Lambda ^{\prime }x+a^{\prime }=\Lambda ^{\prime
}\left( \Lambda x+a\right) +a^{\prime }=\Lambda ^{\prime }\Lambda x+\left(
\Lambda ^{\prime }a+a^{\prime }\right) .
\end{align*}
But $x^{\prime \prime }=g\left( a^{\prime },\Lambda ^{\prime }\right)
x^{\prime }=g\left( a^{\prime },\Lambda ^{\prime }\right) g\left( a,\Lambda
\right) x\equiv g\left( a^{\prime \prime },\Lambda ^{\prime \prime }\right)
x,$ which implies
\begin{equation*}
g\left( a^{\prime },\Lambda ^{\prime }\right) g\left( a,\Lambda \right)
=g\left( \Lambda ^{\prime }a+a^{\prime },\Lambda ^{\prime }\Lambda \right) .
\end{equation*}
The subgroup $SO(3,1)$ is defined by $g\left( 0,\Lambda ^{\prime }\right)
g\left( 0,\Lambda \right) =g\left( 0,\Lambda ^{\prime }\Lambda \right) $ and
the abelian subgroup $T^{4}$ by $g\left( a^{\prime },1\right) g\left(
a,1\right) =g\left( a+a^{\prime },1\right) .$ The inverse transformation is
determined by the two conditions $\Lambda ^{\prime }a+a^{\prime }=0,$ and $%
\Lambda ^{\prime }\Lambda =1$ which implies
\begin{equation*}
g^{-1}\left( a,\Lambda \right) =g\left( -\Lambda ^{-1}a,\Lambda ^{-1}\right)
.
\end{equation*}

\section{Poincare Lie algebra}

We write $\Lambda _{\,\,\nu }^{\mu }=\delta _{\nu }^{\mu }+\zeta _{\,\,\nu
}^{\mu }$ where $\zeta _{\,\,\nu }^{\mu }$ is infinitesimal. This equation
implies that $\zeta _{\mu \nu }=\eta _{\mu \lambda }\zeta _{\,\,\nu
}^{\lambda }=-\zeta _{\nu \mu }$ is antisymmetric. Assume $a^{\mu }$ is
infinitesimal then we need four essential parameters for translation and six
essential parameters for rotations
\begin{equation*}
U(a,\Lambda )=1+i\delta a^{\mu }P_{\mu }-\frac{i}{2}\zeta ^{\mu \nu }J_{\mu
\nu }.
\end{equation*}
The commutation relations of the Lie algebra can be obtained via the usual
basis of scalar functions
\begin{equation*}
U(a,\Lambda )f(x)=f\left( U^{-1}\left( a,\Lambda \right) x\right) =f\left(
\Lambda ^{-1}\left( x-a\right) \right) .
\end{equation*}
By going to the infinitesimal limit we get
\begin{align*}
\left( 1+i\delta a^{\mu }P_{\mu }-\frac{i}{2}\zeta ^{\mu \nu }J_{\mu \nu
}\right) f\left( x^{\lambda }\right) & =f\left( x^{\lambda }-\delta
a^{\lambda }-\zeta _{\,\,\kappa \,}^{\lambda }x^{\kappa }\right)  \\
& =f\left( x^{\lambda }\right) -\left( \delta a^{\lambda }+\zeta
_{\,\,\kappa \,}^{\lambda }x^{\kappa }\right) \partial _{\lambda }f\left(
x\right) +\cdots
\end{align*}
This implies that
\begin{align*}
P_{\mu }& =i\frac{\partial }{\partial x^{\mu }}\equiv i\partial _{\mu }, \\
J_{\mu \nu }& =\left( x_{\mu }P_{\nu }-x_{\nu }P_{\mu }\right) .
\end{align*}
These generators obey the Lie algegra
\begin{align*}
\left[ P_{\mu },P_{\nu }\right] & =0, \\
\left[ J_{\mu \nu },P_{\lambda }\right] & =i\left( \eta _{\nu \lambda
}P_{\mu }-\eta _{\mu \lambda }P_{\nu }\right) , \\
\left[ J_{\mu \nu },J_{\rho \sigma }\right] & =-i\left( \eta _{\mu \rho
}J_{\nu \sigma }-\eta _{\nu \rho }J_{\mu \sigma }-\eta _{\mu \sigma }J_{\nu
\rho }+\eta _{\nu \sigma }J_{\mu \rho }\right) .
\end{align*}
By exponntiating, a general finite transformation is expressible in the form
$U\left( a,\Lambda \right) =\exp \left( i\delta a^{\mu }P_{\mu }-\frac{i}{2}%
\zeta ^{\mu \nu }J_{\mu \nu }\right) $. The Casimir invariants (operators
that commute with the generators of the Lie algebra) are $P^{2}=P^{\mu }\eta
_{\mu \nu }P^{\nu }$ and $W^{2}=W^{\mu }\eta _{\mu \nu }W^{\nu }$ where $%
W_{\mu }=-\frac{1}{2}\epsilon _{\mu \nu \rho \sigma }P^{\nu }J^{\rho \sigma
}.$ One can easily verify that
\begin{align*}
\left[ P^{2},P_{\mu }\right] & =0,\quad \left[ P^{2},J_{\mu \nu }\right] =0,
\\
\left[ W^{2},P_{\mu }\right] & =0,\quad \left[ W^{2},J_{\mu \nu }\right] =0.
\end{align*}
If $P^{2}\neq 0$ we can use the spin pseudo-vector $S_{\mu }=\frac{1}{\sqrt{%
P^{2}}}W_{\mu }$. The Casimir eigenvalues of $P^{2}$ and $W^{2}$ (or $%
j\left( j+1\right) $ associated with $S^{2}$) are used to specify
irreducible representations:
\begin{align*}
P_{\mu }\left| p,j,\lambda \right\rangle & =p_{\mu }\left| p,j,\lambda
\right\rangle , \\
W^{2}\left| p,j,\lambda \right\rangle & =W^{2}\left( j\right) \left|
p,j,\lambda \right\rangle , \\
W_{0}\left| p,j,\lambda \right\rangle & =W_{0}\left( j\right) \left|
p,j,\lambda \right\rangle .
\end{align*}
There are four distinct cases. i) $P^{2}>0,$ ii) $P^{2}=0,$ iii) $P^{2}<0,$
iv) $P_{\mu }=0.$ If $P^{2}\neq 0$ then we can take $P_{1}=P_{2}=0.$ The
little group leaving this choice invariant contains $J_{3}=J_{12},$ so we
can associate $\lambda $ with integer or $\frac{1}{2}$-integer eigenvalues
of $J_{3}:$%
\begin{equation*}
J_{3}\left| \widehat{p},j,\lambda \right\rangle =\lambda \left| \widehat{p}%
,j,\lambda \right\rangle ,
\end{equation*}
so that $\lambda $ is always discrete.

If $P^{2}=m^{2}>0$ we choose the rest frame
\begin{align*}
P_{\mu }& =\pm \left( m,0,0,0\right) , \\
W_{\mu }& =\pm m\left( 0,\overrightarrow{J}\right) ,\quad J_{ij}=\epsilon
_{ijk}J^{k}, \\
S_{\mu }& =\pm \left( 0,\overrightarrow{J}\right) .
\end{align*}
The little group is $O(3)$ generated by $\left[ J_{i},J_{j}\right] =\epsilon
_{ijk}J^{k}$ and is described by the state vectors
\begin{equation*}
\left| j,\lambda \right\rangle ,\quad \lambda =-j,\cdots ,j,\quad j=0,\frac{1%
}{2},1,\cdots
\end{equation*}
where $W^{2}=-m^{2}j(j+1).$

If $P^{2}=0,$ we can take $P_{\mu }=(w,0,0,w),$ and $W_{\mu
}=w(J_{3},L_{1},L_{2},J_{3})$ where $L_{1}=J_{1}+K_{2},$ $L_{2}=J_{2}-K_{1}$
and $J_{0i}=K_{0i},\quad i=1,2,3.$ The generators $L_{1}$ $L_{2}$ and $J_{3}$
satisfy the commutation relations
\begin{equation*}
\left[ L_{1},L_{2}\right] =0,\quad \left[ J_{3},L_{1}\right] =-iL_{2},\quad %
\left[ J_{3},L_{2}\right] =-iL_{1}.
\end{equation*}
The relations $P^{2}=0,$ $W^{2}=0$ and $P^{\mu }W_{\mu }=0$ imply that $%
P^{\mu }$ and $W^{\mu }$ are parallel vectors and therefore we can write
\begin{equation*}
W_{\mu }=\lambda P^{\mu },
\end{equation*}
where $\lambda $ is the helicity. This implies that $W^{2}=-w^{2}L^{2}$ and
unitary representations are labelled by $\left| l,\lambda \right\rangle .$

For the Lorentz group we can write
\begin{equation*}
N_{i}=\frac{1}{2}(J_{i}+iK_{i}),\quad N_{i}^{\dagger }=\frac{1}{2}%
(J_{i}-iK_{i}),
\end{equation*}
where these generators satisfy the commutation relations
\begin{align*}
\left[ N_{i},N_{j}^{\dagger }\right] & =0, \\
\left[ N_{i},N_{j}\right] & =i\epsilon _{ijk}N^{k}, \\
\left[ N_{i}^{\dagger },N_{j}^{\dagger }\right] & =i\epsilon
_{ijk}N^{\dagger k}.
\end{align*}
The Casimir operators are given, in terms of the generators $N_{i}$ and $%
N_{i}^{\dagger },$ by $N_{i}N_{i}$ and $N_{i}^{\dagger }N_{i}^{\dagger }$
with eigenvalues $n(n+1)$ and $m(m+1)$ respectively. The $N_{i}$ and $%
N_{i}^{\dagger }$ obey independently Lie algebras of $SU(2).$ We can label
representations with $(m,n)$ where $m,n=0,\frac{1}{2},1,\frac{3}{2},\cdots $
are the separate $SU(2)$ numbers. The two $SU(2)$'s are not independent. The
spin representation are given by $m+n.$ As examples we have:

\begin{itemize}
\item  $(0,0)$ is spin zero and is the scalar representation.

\item  $(\frac{1}{2},0)$ or $(0,\frac{1}{2})$ is the spin-$\frac{1}{2}$
representation for Weyl spinors, while Dirac spinors are represented by $(%
\frac{1}{2},0)\oplus(0,\frac{1}{2}).$

\item  $(\frac{1}{2},\frac{1}{2})=(\frac{1}{2},0)\otimes(0,\frac{1}{2})$ is
the spin-$1$ representation.
\end{itemize}

We represent $(\frac{1}{2},0)$ and $(0,\frac{1}{2})$ by two-component
complex spinors $\psi _{\alpha }$ and $\overline{\psi }^{\overset{.}{\alpha }%
}$ which transform under Lorentz transformation as
\begin{equation*}
\psi _{\alpha }^{\prime }=M_{\alpha }^{\,\,\beta }\psi _{\beta },\quad
\overline{\psi }^{\overset{.}{\alpha }}=\left( M^{\ast -1}\right) _{\overset{%
.}{\beta }}^{\overset{.}{\,\,\alpha }}\overline{\psi }^{\overset{.}{\beta }}
\end{equation*}
This can be denoted by $\psi \rightarrow M\psi $ and $\overline{\psi }%
\rightarrow \left( M^{\dagger }\right) ^{-1}\overline{\psi }.$ We identify
generators $J_{i}$ with $\frac{1}{2}\sigma _{i}$ and $K_{i}$ with $-\frac{i}{%
2}\sigma _{i}.$ We can therefore write
\begin{equation*}
M=e^{\frac{i}{2}\sigma _{i}\left( w^{i}-iv^{i}\right) },\quad \left(
M^{\dagger }\right) ^{-1}=e^{\frac{i}{2}\sigma _{i}\left(
w^{i}+iv^{i}\right) }.
\end{equation*}
We also introduce the $2\times 2$ matrices $\left( \sigma ^{\mu }\right)
_{\alpha \overset{.}{\beta }}$ defined \ by $\left( \sigma ^{0}\right)
_{\alpha \overset{.}{\beta }}=I_{2},$ the identity matrix and $\left( \sigma
^{i}\right) _{\alpha \overset{.}{\beta }}=-\sigma _{i}$ are the Pauli
matrices. The matrix $P$ is defined by $P=P_{\mu }\left( \sigma ^{\mu
}\right) $ and transforms under Lorentz transformations according to $%
P^{\prime }=MPM^{\dagger }$ or equivalently $P_{\mu }\sigma ^{\mu
}\rightarrow P_{\mu }M\sigma ^{\mu }M^{\dagger }.$ We also define
\begin{equation*}
\overline{\sigma }^{\mu \overset{.}{\alpha }\beta }=\epsilon ^{\overset{.}{%
\alpha }\overset{.}{\gamma }}\epsilon ^{\beta \delta }\sigma _{\,\,\delta
\overset{.}{\gamma }}^{\mu }.
\end{equation*}
These satisfy the proporties
\begin{align*}
\left( \sigma ^{\mu }\overline{\sigma }^{\nu }+\sigma ^{\nu }\overline{%
\sigma }^{\mu }\right) _{\alpha }^{\,\,\beta }& =2\eta ^{\mu \nu }\delta
_{\alpha }^{\beta }, \\
\left( \overline{\sigma }^{\mu }\sigma ^{\nu }+\overline{\sigma }^{\nu
}\sigma ^{\mu }\right) _{\overset{.}{\alpha }}^{\overset{.}{\,\,\beta }}&
=-2\eta ^{\mu \nu }\delta _{\overset{.}{\alpha }}^{\overset{.}{\beta }}.
\end{align*}
From these matrices we can construct representations of the Lorentz algebra.
They are given by
\begin{align*}
\sigma _{\quad \alpha }^{\mu \nu \,\,\beta }& =\left( \sigma ^{\mu }%
\overline{\sigma }^{\nu }-\sigma ^{\nu }\overline{\sigma }^{\mu }\right)
_{\alpha }^{\,\,\beta }, \\
\overline{\sigma }_{\qquad \overset{.}{\alpha }}^{\mu \nu \,\,\overset{.}{%
\beta }}& =\left( \overline{\sigma }^{\mu }\sigma ^{\nu }-\overline{\sigma }%
^{\nu }\sigma ^{\mu }\right) _{\overset{.}{\alpha }}^{\overset{.}{\,\,\beta }%
}.
\end{align*}
A Dirac spinor takes the form
\begin{equation*}
\psi =\left(
\begin{array}{c}
\chi _{\alpha } \\
\overline{\lambda }^{\overset{.}{\alpha }}
\end{array}
\right) ,
\end{equation*}
and the Dirac gamma matrices are defined by
\begin{equation*}
\gamma ^{\mu }=\left(
\begin{array}{cc}
0 & \sigma ^{\mu } \\
\overline{\sigma }^{\mu } & 0
\end{array}
\right) ,
\end{equation*}
which satisfy the anticommutation relations $\left\{ \gamma ^{\mu },\gamma
^{\nu }\right\} =2\eta ^{\mu \nu }I_{4}.$ This defines a Clifford algebra.
The representation of the Lorentz algebra in terms of the gamma matrices is
given by
\begin{equation*}
S^{\mu \nu }=\frac{i}{4}\left[ \gamma ^{\mu },\gamma ^{\nu }\right] .
\end{equation*}
Let $\Lambda _{\frac{1}{2}}=\exp \left( -\frac{i}{2}\zeta _{\mu \nu }S^{\mu
\nu }\right) $ then $\Lambda _{\frac{1}{2}}^{-1}\gamma ^{\mu }\Lambda _{%
\frac{1}{2}}=\Lambda _{\quad \nu }^{\mu }\gamma ^{\nu }$ where $\Lambda
_{\quad \nu }^{\mu }=\delta _{\mu }^{\nu }+\zeta _{\quad \nu }^{\mu }.$
These proporties will be used in the next section when we derive the Dirac
equation.

\section{Dynamical equations for free and interacting fields}

\subsection{Spin-$0:$ Klein-Gordon field}

The Hamiltonian in relativistic dynamics has eigenvalues $E$ which can be
determined from the relation $P_{\mu }P^{\mu }=m^{2}$. This relation implies
that $P_{0}^{2}=\overrightarrow{P}.\overrightarrow{P}+m^{2}=E^{2}$, where we
have set the velocity of light $c$ to $1$. Using the quantum mechanical
correspondence $P_{\mu }=\frac{\hbar }{i}\frac{\partial }{\partial x^{\mu }}$
and setting $\hbar =1$) and acting with the operator $\left( P_{\mu }P^{\mu
}-m^{2}\right) $ on the scalar field $\phi $ gives
\begin{equation*}
\left( \eta ^{\mu \nu }\partial _{\mu }\partial _{\nu }+m^{2}\right) \phi =0,
\end{equation*}
which is the Klein-Gordon equation. This equation could be derived from a
varriational principle using the action
\begin{equation*}
I_{0}=\int d^{4}xL=\frac{1}{2}\int d^{4}x\left( \eta ^{\mu \nu }\partial
_{\mu }\phi \partial _{\nu }\phi -m^{2}\phi ^{2}\right) .
\end{equation*}
The Euler-Lagrange equation obtained by demanding that $\delta I=0,$ with
the field $\phi $ vanishing at the boundary, is given by
\begin{equation*}
\frac{\delta L}{\delta \phi }=\partial _{\mu }\left( \frac{\delta L}{\delta
\partial _{\mu }\phi }\right) ,
\end{equation*}
and this implies the Klein-Gordon equation.

\subsection{Spin-$\frac{1}{2}:$ Dirac field}

The Schr\"{o}dinger equation is non-relativistic and is linear in
time-derivatives but quadratic in space-derivatives
\begin{equation*}
i\frac{\partial \psi }{\partial t}=-\frac{{\hbar }^{2}}{2m}\overrightarrow{%
\nabla }^{2}\psi .
\end{equation*}
The relativistic equation should be linear in time and one must find a first
order differential equation of the form
\begin{equation*}
i\hbar \frac{\partial \psi }{\partial t}=\frac{\hbar c}{i}\alpha ^{i}\frac{%
\partial \psi }{\partial x^{i}}+\beta m\psi ,
\end{equation*}
where $\alpha ^{i}$ and $\beta $ are $4\times 4$ matrices. The square of
this equation should yield an equation of the Klein-Gordon type. This is
possible if $\alpha ^{i}$ and $\beta $ satisfy the proporties
\begin{equation*}
\alpha ^{i}\alpha ^{j}+\alpha ^{j}\alpha ^{i}=2\delta ^{ij},\quad \beta
^{2}=1,\quad \alpha ^{i}\beta +\beta \alpha ^{i}=0.
\end{equation*}
Multiplying both sides of the equation by $\beta $ and defining $\gamma
^{0}=\beta ,$ $\gamma ^{i}=\beta \alpha ^{i}$ gives
\begin{equation*}
\left( i\gamma ^{\mu }\partial _{\mu }-m\right) \psi =0.
\end{equation*}
The matrices $\gamma ^{\mu }$ satisfy the anticommutation relations $\left\{
\gamma ^{\mu },\gamma ^{\nu }\right\} =2\eta ^{\mu \nu }$ and can be
identified with the gamma matrices defined in the previous section. This is
the Dirac equation. One can easily verify that it is Lorentz covariant
\begin{align*}
\left( i\gamma ^{\mu }\partial _{\mu }-m\right) \psi & \rightarrow \left(
i\gamma ^{\mu }\Lambda _{\mu }^{-1\nu }\partial _{\nu }-m\right) \Lambda _{%
\frac{1}{2}}\psi \left( \Lambda ^{-1}x\right)  \\
& =\Lambda _{\frac{1}{2}}\Lambda _{\frac{1}{2}}^{-1}\left( i\gamma ^{\mu
}\Lambda _{\mu }^{-1\nu }\partial _{\nu }-m\right) \Lambda _{\frac{1}{2}%
}\psi \left( \Lambda ^{-1}x\right)  \\
& =\Lambda _{\frac{1}{2}}\left( i\gamma ^{\mu }\partial _{\mu }-m\right)
\psi \left( \Lambda ^{-1}x\right) ,
\end{align*}
where we have used the property $\Lambda _{\frac{1}{2}}^{-1}\gamma ^{\mu
}\Lambda _{\frac{1}{2}}=\Lambda _{\,\,\nu }^{\mu }\gamma ^{\nu }$ derived in
the last section.

The action
\begin{equation*}
I_{\frac{1}{2}}=\int d^{4}x\overline{\psi }(x)\left( i\gamma ^{\mu }\partial
_{\mu }-m\right) \psi (x),
\end{equation*}
is Lorentz invariant because $\overline{\psi }\rightarrow \overline{\psi }%
\Lambda _{\frac{1}{2}}^{-1}.$ In the special case when $m=0,$ the Dirac
equation simplifies to
\begin{equation*}
\left(
\begin{array}{cc}
0 & i\sigma ^{\mu }\partial _{\mu } \\
i\overline{\sigma }^{\mu }\partial _{\mu } & 0
\end{array}
\right) \left(
\begin{array}{c}
\psi _{L} \\
\psi _{R}
\end{array}
\right) =0,
\end{equation*}
and this splits into two independent equations, one for the left-handed
component $\psi _{L}$ and the other for the right handed component $\psi _{R}
$%
\begin{equation*}
i\sigma ^{\mu }\partial _{\mu }\psi _{L}=0,\quad i\overline{\sigma }^{\mu
}\partial _{\mu }\psi _{R}=0.
\end{equation*}

\subsection{Spin-$1$: Maxwell field}

We are interested in  massless vectors, in particular, the photon which
obeys Maxwell's equations
\begin{align*}
\overrightarrow{\nabla }\cdot \overrightarrow{E}& =\rho ,\quad
\overrightarrow{\nabla }\times \overrightarrow{E}=-\frac{\partial
\overrightarrow{B}}{\partial t}, \\
\overrightarrow{\nabla }\cdot \overrightarrow{B}& =0,\quad \overrightarrow{%
\nabla }\times \overrightarrow{B}=\frac{\partial \overrightarrow{E}}{%
\partial t}+\overrightarrow{J}.
\end{align*}
The second and fourth equations are solved by
\begin{equation*}
\overrightarrow{E}=-\overrightarrow{\nabla }A^{0}-\frac{\partial
\overrightarrow{A}}{\partial t},\quad \overrightarrow{B}=\overrightarrow{%
\nabla }\times \overrightarrow{A},
\end{equation*}
and are invariant under the gauge transformations
\begin{equation*}
\overrightarrow{A}^{\prime }=\overrightarrow{A}-\overrightarrow{\nabla }%
\Lambda ,\quad A_{0}^{\prime }=A_{0}+\frac{\partial \Lambda }{\partial t}.
\end{equation*}
Therefore one can impose a gauge choice, the Lorentz gauge, on the potential
$\overrightarrow{A}$ and $A_{0}$%
\begin{equation*}
\nabla \cdot A+\frac{\partial A^{0}}{\partial t}=0.
\end{equation*}
In this gauge, Maxwell equations simplify to
\begin{equation*}
\eta ^{\mu \nu }\partial _{\mu }\partial _{\nu }A^{0}=-\rho ,\quad \eta
^{\mu \nu }\partial _{\mu }\partial _{\nu }\overrightarrow{A}=%
\overrightarrow{J},
\end{equation*}
which is a standard wave equation with a source. Maxwell equations could be
written in a Lorentz covariant manner if we define the curvature
\begin{equation*}
F_{\mu \nu }=\partial _{\mu }A_{\nu }-\partial _{\nu }A_{\mu }
\end{equation*}
so that
\begin{equation*}
\partial _{\nu }F^{\mu \nu }=J^{\nu }.
\end{equation*}
We can replace the definition of $F_{\mu \nu }$ by the requirement
\begin{equation*}
\partial _{\rho }F_{\mu \nu }+\partial _{\mu }F_{\nu \rho }+\partial _{\nu
}F_{\rho \mu }=0,
\end{equation*}
which, locally, can be solved by the above equation defining $A_{\mu }.$ The
curvature $F_{\mu \nu }$ is invariant under the gauge transformations $%
A_{\mu }^{\prime }=A_{\mu }+\partial _{\mu }\Lambda .$ Maxwell equations
could be obtained from the action
\begin{equation*}
I_{1}=\int d^{4}x\left( -\frac{1}{4}F_{\mu \nu }F_{\rho \sigma }\eta ^{\mu
\rho }\eta ^{\nu \sigma }+A_{\mu }J^{\mu }\right) .
\end{equation*}

According to Noether's theorem, there is a conserved charged associated to
every symmetry of the action. The symmetry in this case gives electric
charge conservation. To find the interactions of  electrons with photons, we
note that the action for electrons $I_{\frac{1}{2}}$ is invariant under the
global symmetry $\psi (x)\rightarrow e^{-ie\Lambda }\psi (x)$ where $\Lambda
$ is a constant phase. This global symmetry is lost if the symmetry is
promoted to a local symmetry by allowing $\Lambda $ to depend on $x.$ In
this case
\begin{equation*}
\overline{\psi }(x)i\gamma ^{\mu }\partial _{\mu }\psi (x)\rightarrow
\overline{\psi }(x)i\gamma ^{\mu }\partial _{\mu }\psi (x)+e\partial _{\mu
}\Lambda \overline{\psi }(x)\gamma ^{\mu }\psi (x).
\end{equation*}
To restore the symmetry, we add to the Dirac-action, the photon-electron
interaction term
\begin{equation*}
-e\int d^{4}x\overline{\psi }(x)\gamma ^{\mu }\psi (x)A_{\mu },
\end{equation*}
which transforms under the local symmetry according to
\begin{equation*}
-e\overline{\psi }(x)\gamma ^{\mu }\psi (x)A_{\mu }\rightarrow -e\overline{%
\psi }(x)\gamma ^{\mu }\psi (x)A_{\mu }-e\partial _{\mu }\Lambda \overline{%
\psi }(x)\gamma ^{\mu }\psi (x).
\end{equation*}
We deduce that the action for the spin-$\frac{1}{2},$ spin-$1$ system
governing the interaction of photons and electrons and invariant under the
local gauge transformations $\psi (x)\rightarrow e^{-ie\Lambda }\psi (x)$
and $A_{\mu }\rightarrow A_{\mu }+\partial _{\mu }\Lambda $, is given by
\begin{equation*}
I_{\left( \frac{1}{2},1\right) }=\int d^{4}x\left( -\frac{1}{4}F_{\mu \nu
}F^{\mu \nu }+\overline{\psi }(x)\left( i\gamma ^{\mu }D_{\mu }-m\right)
\psi (x)\right) ,
\end{equation*}
where $D_{\mu }=\partial _{\mu }+ieA_{\mu }.$ Therefore, interaction terms
are obtained by replacing ordinary derivatives $\partial _{\mu }$ in the
free Dirac action by covariant derivatives $D_{\mu }.$ These covariant
derivatives satisfy the commutation relations
\begin{equation*}
\left[ D_{\mu },D_{\nu }\right] =ieF_{\mu \nu }.
\end{equation*}

\subsection{Spin-$\frac{3}{2}:$ Rarita-Schwinger field}

The field that  describes a spin-$\frac{3}{2}$ particle is a vector-spinor $%
\psi _{\mu \alpha }$ where $\mu $ is a space-time vector index and $\alpha $
is a spinor index. The free field satisfies the Rarita-Schwinger equation
\begin{equation*}
\gamma ^{\mu \nu \rho }\psi _{\nu \rho }=0
\end{equation*}
where $\psi _{\mu \nu }=\partial _{\mu }\psi _{\nu }-\partial _{\nu }\psi
_{\mu },$ and
\begin{equation*}
\gamma ^{\mu \nu \rho }=\frac{1}{3!}\left( \gamma ^{\mu }\gamma ^{\nu
}\gamma ^{\rho }-\gamma ^{\nu }\gamma ^{\mu }\gamma ^{\rho }+\gamma ^{\nu
}\gamma ^{\rho }\gamma ^{\mu }-\gamma ^{\rho }\gamma ^{\nu }\gamma ^{\mu
}+\gamma ^{\rho }\gamma ^{\mu }\gamma ^{\nu }-\gamma ^{\mu }\gamma ^{\rho
}\gamma ^{\nu }\right)
\end{equation*}
is completely antisymmetric in the indices $\mu \nu \rho .$ The product of
spin-$1$ and spin-$\frac{1}{2}$ representations gives spin-$\frac{3}{2}$ and
spin-$\frac{1}{2}$ representations
\begin{equation*}
\left( \frac{1}{2},\frac{1}{2}\right) \otimes \left( \frac{1}{2},0\right)
=\left( 1,\frac{1}{2}\right) \oplus \left( 0,\frac{1}{2}\right) .
\end{equation*}
It will be seen that the Rarita-Schwinger equation is necessary to eliminate
the spin-$\frac{1}{2}$ component from $\psi _{\mu \alpha }.$

\subsection{Spin-$2:$ Gravitational field}

The massless spin-$2$ field is the graviton which is a symmetric traceless
tensor $h_{\mu \nu }$ with gauge invariance
\begin{equation*}
h_{\mu \nu }\rightarrow h_{\mu \nu }+\partial _{\mu }\xi _{\nu }+\partial
_{\nu }\xi _{\mu }.
\end{equation*}
To see that this is none other than the linearized form of the metric tensor
of the underlying space-time manifold, let us assume that we would like to
promote the global Lorentz invariance of spinors to a local one:
\begin{equation*}
\psi _{\alpha }\left( x\right) \rightarrow \exp \left( \frac{i}{4}\Lambda
^{ab}\left( x\right) \gamma _{ab}\right) _{\alpha }^{\beta }\psi _{\beta
}\left( x\right) ,
\end{equation*}
where $\gamma _{ab}=\frac{1}{2}\left( \gamma ^{a}\gamma ^{b}-\gamma
^{b}\gamma ^{a}\right) .$ To restore the invariance of the Dirac equation
under local Lorentz transformations, and in analogy with the electromagnetic
case, we introduce the covariant derivative
\begin{equation*}
D_{\mu }=\partial _{\mu }+\frac{1}{4}\omega _{\mu }^{\,\,ab}\gamma _{ab}.
\end{equation*}
The term $\overline{\psi }(x)\left( i\gamma ^{\mu }D_{\mu }-m\right) \psi (x)
$ becomes invariant provided that
\begin{equation*}
\omega _{\mu }^{\prime ab}=\partial _{\mu }\Lambda ^{ab}+\omega _{\mu
}^{\,\,ac}\Lambda _{c}^{\,b}-\omega _{\mu }^{\,\,bc}\Lambda _{c}^{\,a}.
\end{equation*}
The curvature tensor is defined by
\begin{equation*}
\left[ D_{\mu },D_{\nu }\right] =\frac{1}{4}R_{\mu \nu }^{\quad ab}\gamma
_{ab},
\end{equation*}
which can be easily evaluated to be
\begin{equation*}
R_{\mu \nu }^{\quad ab}=\partial _{\mu }\omega _{\nu }^{\,\,ab}-\partial
_{\nu }\omega _{\mu }^{\,\,ab}+\omega _{\mu }^{\,\,ac}\omega _{\nu c}^{\quad
b}-\omega _{\nu }^{\,\,ac}\omega _{\mu c}^{\quad b},
\end{equation*}
and is covariant under local Lorentz transformations. To make contact with
the Riemann curvature tensor, we introduce the soldering form $e_{\mu }^{a}$
satisfying
\begin{equation*}
\nabla _{\mu }e_{\nu }^{a}=\partial _{\mu }e_{\nu }^{a}-\Gamma _{\mu \nu
}^{\rho }e_{\rho }^{a}+\omega _{\mu }^{\,\,ab}e_{\nu b}=0,
\end{equation*}
where $\Gamma _{\mu \nu }^{\rho }=\frac{1}{2}g^{\rho \sigma }\left( g_{\mu
\sigma ,\nu }+g_{\nu \sigma ,\mu }-g_{\mu \nu ,\sigma }\right) $ is the
Christoffel connection and $g_{\mu \nu }=e_{\mu }^{a}\eta _{ab}e_{\nu }^{b}.$
This equation implies that the covariant derivative of the metric under
coordinate transformations vanishes. The constraint equation can be solved
for $\omega _{\mu }^{\,\,ab}$ in terms of $e_{\mu }^{a}.$ Substituting the
solution into $R_{\mu \nu }^{\quad ab}\left( \omega \right) $ one can show
that
\begin{equation*}
R_{\mu \nu }^{\quad ab}\left( \omega \right) e_{\rho a}e_{\sigma b}=R_{\mu
\nu \rho \sigma }(g),
\end{equation*}
is identical to the Riemann tensor as a function of the metric $g.$

Equivalently, we can introduce gauge fields corresponding to both the
Lorentz generators $J_{ab}$ and translations $P_{a}$%
\begin{equation*}
D_{\mu }=\partial _{\mu }+\omega _{\mu }^{\,\,ab}J_{ab}+e_{\mu }^{a}P_{a},
\end{equation*}
then the curvature is
\begin{equation*}
\left[ D_{\mu },D_{\nu }\right] =R_{\mu \nu }^{\quad ab}J_{ab}+T_{\mu \nu
}^{a}P_{a},
\end{equation*}
where $R_{\mu \nu }^{\quad ab}$ is the same as before while
\begin{equation*}
T_{\mu \nu }^{a}=\partial _{\mu }e_{\nu }^{a}-\partial _{\nu }e_{\mu
}^{a}+\omega _{\mu }^{\,\,ab}e_{\nu b}-\omega _{\nu }^{\,\,ab}e_{\mu b}.
\end{equation*}
By setting the torsion $T_{\mu \nu }^{a}$to zero we can uniquely solve for $%
\omega _{\mu }^{\,\,ab}$ in terms of $e_{\mu }^{a}$ and substitute this
expression into $R_{\mu \nu }^{\quad ab}$ to obtain the Riemann tensor.

This last formulation is related to the Cartan structure equations
\begin{align*}
T^{a}& =de^{a}+\omega ^{ab}\wedge e_{b}, \\
R^{ab}& =d\omega ^{ab}+\omega ^{ac}\wedge \omega _{c}^{\,\,b}.
\end{align*}
By writing $e^{a}=e_{\mu }^{a}dx^{\mu },$ $\omega ^{ab}=\omega _{\mu
}^{ab}dx^{\mu },$ $R^{ab}=\frac{1}{2}R_{\mu \nu }^{\quad ab}dx^{\mu }\wedge
dx^{\nu }$ and $T^{a}=\frac{1}{2}T_{\mu \nu }^{a}dx^{\mu }\wedge dx^{\nu }$
we can recover all the previous formulas.

The Einstein-Hilbert action is given by
\begin{equation*}
I_{2}=-\int d^{4}x\frac{e}{4}e_{a}^{\mu }e_{b}^{\nu }R_{\mu \nu }^{\quad ab},
\end{equation*}
where $e=\det \left( e_{\mu }^{a}\right) .$ The dynamical equations
governing the interaction of electrons and photons with gravity can be read
from the equation
\begin{align*}
I_{\left( \frac{1}{2},1,2\right) }& =\int d^{4}xe\left( \overline{\psi }%
\left( i\gamma ^{\mu }\left( \partial _{\mu }+eA_{\mu }+\frac{1}{4}\omega
_{\mu }^{\,\,ab}\gamma _{ab}\right) -m\right) \psi \right)  \\
& -\int d^{4}xe\left( \frac{1}{4}F_{\mu \nu }F_{\rho \sigma }g^{\mu \rho
}g^{\nu \sigma }+\frac{1}{4}e_{a}^{\mu }e_{b}^{\nu }R_{\mu \nu }^{\quad
ab}\right) .
\end{align*}

\subsection{The standard model}

The principle of gauge invariance plays a fundamental role in physics. The
Poincare symmetry is a space-time symmetry effecting local space-time
coordinates. On the other hand the known elementary particles in nature,
fall into group representations, e.g. fermions have the local gauge
invariance under the transformations $\psi _{\alpha I}\rightarrow \left(
e^{ig_{a}\Lambda ^{a}\left( x\right) T^{a}}\right) _{I}^{J}\psi _{\alpha J}$
where $\left( T^{a}\right) _{I}^{J}$ are the representations of the gauge
group. As in the Maxwell case, connections are introduced to insure local
gauge invariance. Let
\begin{equation*}
D=d+igA,\quad A=A_{\mu }^{a}T^{a}dx^{\mu },
\end{equation*}
so that the curvature of the connection $A$ will be given by
\begin{align*}
D^{2}& =igF=ig\left( dA+igA^{2}\right) , \\
F& =\frac{1}{2}F_{\mu \nu }^{a}T^{a}dx^{\mu }\wedge dx^{\nu },
\end{align*}
so that
\begin{equation*}
F_{\mu \nu }^{a}=\partial _{\mu }A_{\nu }^{a}-\partial _{\nu }A_{\mu
}^{a}+gf^{abc}A_{\mu }^{b}A_{\nu }^{c},
\end{equation*}
where $\left[ T^{a},T^{b}\right] =if^{abc}T^{c}.$ Let us denote the quarks
by $q=\left(
\begin{array}{c}
u_{L} \\
d_{L}
\end{array}
\right) ,$ $u_{R},$ $d_{R}$ which are in the representations $\left( 3,2,-%
\frac{1}{3}\right) ,$ $\left( \overline{3},1,-\frac{4}{3}\right) $ and $(%
\overline{3},1,\frac{2}{3})$ of $SU(3)_{c}\times SU(2)_{w}\times U(1)_{Y}.$
The leptons are denoted by $l=\left(
\begin{array}{c}
e_{L}^{-} \\
\nu _{L}
\end{array}
\right) ,$ $e_{_{R}^{+}}$ which are in the $\left( 1,2,1\right) $ and $%
\left( 1,1,2\right) $ of $SU(3)_{c}\times SU(2)_{w}\times U(1)_{Y}..$ The
Lagrangian that govern all known interactions is given by
\begin{align*}
e^{-1}L& =-\frac{1}{4\kappa ^{2}}R-\frac{1}{4}\left( F_{\mu \nu }^{i}F^{\mu
\nu i}+F_{\mu \nu }^{\alpha }F^{\mu \nu \alpha }+B_{\mu \nu }B^{\mu \nu
}\right)  \\
& +D_{\mu }H^{\dagger }D^{\mu }H-\mu ^{2}H^{\dagger }H+\lambda \left(
H^{\dagger }H\right) ^{2} \\
& +i\overline{q}\gamma ^{\mu }\left( \partial _{\mu }+\frac{1}{4}\omega
_{\mu }^{\,\,ab}\gamma _{ab}-\frac{i}{2}g_{2}A_{\mu }^{\alpha }\sigma
^{\alpha }-\frac{i}{6}g_{1}B_{\mu }-\frac{i}{2}g_{3}V_{\mu }^{i}\lambda
^{i}\right) q \\
& +i\overline{d}_{R}\gamma ^{\mu }\left( \partial _{\mu }+\frac{1}{4}\omega
_{\mu }^{\,\,ab}\gamma _{ab}+\frac{i}{3}g_{1}B_{\mu }-\frac{i}{2}g_{3}V_{\mu
}^{i}\lambda ^{i}\right) d_{R} \\
& +i\overline{u}_{R}\gamma ^{\mu }\left( \partial _{\mu }+\frac{1}{4}\omega
_{\mu }^{\,\,ab}\gamma _{ab}-\frac{2i}{3}g_{1}B_{\mu }-\frac{i}{2}%
g_{3}V_{\mu }^{i}\lambda ^{i}\right) u_{R} \\
& +i\overline{l}\gamma ^{\mu }\left( \partial _{\mu }+\frac{1}{4}\omega
_{\mu }^{\,\,ab}\gamma _{ab}-\frac{i}{2}g_{2}A_{\mu }^{\alpha }\sigma
^{\alpha }+\frac{i}{2}g_{1}B_{\mu }\right) l \\
& +i\overline{e}_{R}\gamma ^{\mu }\left( \partial _{\mu }+\frac{1}{4}\omega
_{\mu }^{\,\,ab}\gamma _{ab}+ig_{1}B_{\mu }\right) e_{R} \\
& +\left( k^{d}\overline{q}Hd_{R}+k^{u}\overline{q}\tau ^{2}Hu_{R}+k^{e}%
\overline{l}He_{R}+h.c.\right) ,
\end{align*}
where the Higgs field $H$ is in the $(1,2,1)$ representation so that $D_{\mu
}H=\partial _{\mu }H-\frac{i}{2}g_{2}A_{\mu }^{\alpha }\sigma ^{\alpha }-%
\frac{i}{2}g_{1}B_{\mu }H$ and $V_{\mu }^{i},$ $A_{\mu }^{\alpha }$ and $%
B_{\mu }$ are the gauge fields for the gauge Lie algebras $SU(3)_{c},$ $%
SU(2)_{w}$ and $U(1)_{Y}.$ The $k^{d},$ $k^{u}$ and $k^{e}$ are $3\times 3$
matrices mixing the three generations of particles. Minimizing the potential
of the Higgs fields $H$ gives a vacuum
\begin{equation*}
\left\langle H\right\rangle =\left(
\begin{array}{c}
0 \\
v
\end{array}
\right) ,
\end{equation*}
which generates masses for the quarks and leptons, e.g. the leptons will
acquire the term
\begin{equation*}
k^{e}\overline{l}He_{R}=k^{e}v\overline{e}_{L}e_{R}.
\end{equation*}
Similarly the term $D_{\mu }H^{\dagger }D^{\mu }H$ will yield the mass terms
\begin{equation*}
\frac{1}{4}v^{2}g_{2}^{2}\left( W_{\mu }^{+}W^{\mu -}+\frac{1}{\cos
^{2}\theta }Z_{\mu }Z^{\mu }\right) ,
\end{equation*}
where $W_{\mu }^{\pm }=\left( A_{\mu }\pm iA_{\mu }^{2}\right) ,$ $Z_{\mu }=%
\frac{1}{\sqrt{g_{1}^{2}+g_{2}^{2}}}\left( g_{2}A_{\mu }^{3}-g_{1}B_{\mu
}\right) $ and $\sin ^{2}\theta =\frac{g_{1}^{2}}{g_{1}^{2}+g_{2}^{2}}.$ The
symmetry of the Lagrangian is diffeomorphism$\times $ local internal
symmetry. All fermions are chiral (Weyl fermions) and acquire mass only
after the symmetry is broken spontaneously from $SU(3)_{c}\times
SU(2)_{w}\times U(1)_{Y}$ to $SU(3)_{c}\times U(1)_{em}.$ At the quantum
level chiral fermions break gauge invariance. The chirality condition is $%
\gamma _{5}\psi _{\pm }=\pm \psi _{\pm }$ where $\gamma _{5}=i\gamma
_{0}\gamma _{1}\gamma _{2}\gamma _{3}$ so that $\gamma _{5}^{2}=1.$ A Dirac
spinor is decomposed according to $\psi =\psi _{+}+\psi _{-}.$ The
eigenvalue equation $D\psi =\lambda \psi $ implies $\gamma _{5}D\psi _{\pm
}=\mp \lambda D\psi _{\pm }$ so that the eigenvalue problem could not be set
for chiral fermions alone without their partners except for zero eigevalues.
After quantization one must sum over all states which would involve
computing the determinat
\begin{equation*}
\int D\psi D\overline{\psi }e^{\overline{\psi }D\psi }\rightarrow \det D,
\end{equation*}
and this involve the step of taking the trace over all states, which must
now be restricted to the chiral ones
\begin{equation*}
\left\langle \psi _{n-}\right| D\left| \psi _{n+}\right\rangle .
\end{equation*}
This is not invariant under the chiral rotation $\psi _{n+}\rightarrow
e^{i\theta \gamma _{5}}\psi _{n+}.$ To see this explicitely we have
\begin{equation*}
Tr(F(D))=\sum\limits_{n}\left\langle \psi _{n-}\right| F(D)\left| \psi
_{n+}\right\rangle \rightarrow \sum\limits_{n}\left\langle e^{-i\theta
^{a}T^{a}}\psi _{n-}\right| F(D)\left| e^{i\theta ^{a}T^{a}}\psi
_{n+}\right\rangle .
\end{equation*}
Since $F(D)$ depends only on geometrical quantities, one can use the heat
kernel expansion to find the lowest terms
\begin{align*}
Tr(F(D))& \rightarrow 2i\theta ^{a}Tr\left( \gamma _{5}T^{a}\gamma ^{\mu \nu
}F_{\mu \nu }^{b}T^{b}\gamma ^{\rho \sigma }F_{\rho \sigma }^{c}T^{c}\right)
\\
& =2i\theta ^{a}\epsilon ^{\mu \nu \rho \sigma }F_{\mu \nu }^{b}F_{\rho
\sigma }^{c}Tr\left( T^{a}\left\{ T^{b},T^{c}\right\} \right) .
\end{align*}
We deduce that gauge invariance is not destroyed by chirality if
\begin{equation*}
Tr\left( T^{a}\left\{ T^{b},T^{c}\right\} \right) =0.
\end{equation*}
If we include gravitational terms then the lowest order term is
\begin{equation*}
2i\theta ^{a}Tr\left( \gamma _{5}T^{a}\gamma ^{\mu \nu }R_{\mu \nu }^{\quad
cd}\gamma ^{\rho \sigma }R_{\rho \sigma cd}\right) ,
\end{equation*}
which vanishes if
\begin{equation*}
Tr\left( T^{a}\right) =0,
\end{equation*}
insuring the absence of gravitational anomalies. Both conditions on the
representations of the chiral fermions in the standared model are satisfied.
These are very strong constraints on possible representations of the
fermions in a realistic model and lead to the conclusion that these are not
satisfied by accident  but are derivable from higher principles. One such
explanation is unification where the chiral fermionic representations
\begin{equation*}
\left( 3,2,-\frac{1}{3}\right) \oplus \left( \overline{3},1,\frac{2}{3}%
\right) \oplus \left( \overline{3},1,-\frac{4}{3}\right) \oplus \left(
1,2,1\right) \oplus \left( 1,1,2\right)
\end{equation*}
could be obtained from the $\overline{5}\oplus 10$ representations of $SU(5)$
to be broken spontaneously into two stages
\begin{equation*}
SU(5)\rightarrow SU(3)_{c}\times SU(2)_{w}\times U(1)_{Y}\rightarrow
SU(3)_{c}\times U(1)_{em}.
\end{equation*}
This is not the unique possibility but the simplest.

\section{Supersymmetry}

In nature there are particles of different spins, in particular integral and
half-integral spins. In the ultimate theory there must be a symmetry that
places particles with different spin in the same multiplet. Until 1974 all
internal symmetries were studied and there was a theorem by Coleman and
Mandula stating that, under certain physical assumptions,  it is impossible
to have a space-time symmetry larger than the Poincare symmetry. This
obstruction was overcome by the simple extension of considering $\mathbb{Z}%
_{2}$ graded algebras whose generators are classified into two classes, even
(bosonic) and odd (fermionic) and obey
\begin{equation*}
\left[ even,even\right] =even,\quad \left[ even,odd\right] =odd,\quad
\left\{ odd,odd\right\} =even.
\end{equation*}
With every generator $\ A$ in the graded Lie algebra we associate a number $%
a=\left\{
\begin{array}{c}
0\text{ if }A\text{ is even} \\
1\text{ if }A\text{ is odd}
\end{array}
\right. $ and define the graded commutator by
\begin{equation*}
\left[ A,B\right\} =AB-\left( -1\right) ^{ab}BA,
\end{equation*}
which satisfies the graded Jacobi identity
\begin{equation*}
\left[ A,\left[ B,C\right\} \right\} +(-1)^{a(b+c)}\left[ B,\left[
C,A\right\} \right\} +(-1)^{c(a+b)}\left[ C,\left[ A,B\right\} \right\} =0.
\end{equation*}
Denote by $Q_{\alpha }^{i}$ the fermionic generators where the index $\alpha
$ transforms under the Lorentz group and the index $i$ transforms under an
internal algebra with generators $T_{r}$%
\begin{align*}
\left[ T_{r},T_{s}\right] & =f_{rst}T_{t},\quad \left[ P_{\mu },T_{r}\right]
=0,\quad \left[ J_{\mu \nu },T_{r}\right] =0, \\
\left[ Q_{\alpha }^{i},J_{\mu \nu }\right] & =\left( b_{\mu \nu }\right)
_{\alpha }^{\beta }Q_{\beta }^{i},\quad \left[ Q_{\alpha }^{i},P_{\mu }%
\right] =\left( c_{\mu }\right) _{\alpha }^{\beta }Q_{\beta }^{i}, \\
\left\{ Q_{\alpha }^{i},Q_{\beta }^{j}\right\} & =r\left( \gamma ^{\mu
}C\right) _{\alpha \beta }P_{\mu }\delta ^{ij}+s\left( \gamma ^{\mu \nu
}C\right) _{\alpha \beta }J_{\mu \nu }\delta ^{ij} \\
& +C_{\alpha \beta }U^{ij}+\left( \gamma _{5}C\right) _{\alpha \beta
}V^{ij}+\left( \gamma ^{\mu }\gamma _{5}C\right) _{\alpha \beta }L_{\mu
}^{ij}.
\end{align*}
By the Coleman-Mandula theorem we can set $L_{\mu }^{ij}$ to zero because it
carries a Lorentz vector index. Without any loss in generality we can assume
that $Q_{\alpha }^{i}$ satisfy the Majorana condition $Q_{\alpha
}^{i}=C_{\alpha \beta }\overline{Q}^{\beta i}.$

Using the Jacobi identity
\begin{equation*}
\left[ J_{\mu \nu },\left[ J_{\rho \sigma },Q_{\alpha }^{i}\right] \right] +%
\left[ J_{\rho \sigma },\left[ Q_{\alpha }^{i}\right] \right] +\left[
Q_{\alpha }^{i},\left[ J_{\mu \nu },J_{\rho \sigma }\right] \right] =0,
\end{equation*}
we deduce that $\left( b_{\mu \nu }\right) _{\alpha }^{\beta }$ should form
a representation of the Lorentz group and must be identified with
\begin{equation*}
\left( b_{\mu \nu }\right) _{\alpha }^{\beta }=\frac{i}{2}\left( \gamma
_{\mu \nu }\right) _{\alpha }^{\beta }
\end{equation*}
From the identity
\begin{equation*}
\left[ P_{\mu },\left[ P_{\nu },Q_{\alpha }^{i}\right] \right] +\left[
P_{\nu },\left[ Q_{\alpha }^{i},P_{\mu }\right] \right] +\left[ Q_{\alpha
}^{i},\left[ P_{\mu },P_{\nu }\right] \right] =0,
\end{equation*}
one deduces that $\left[ c_{\mu },c_{\nu }\right] =0.$ But it is always
possible to change $\left( c_{\mu }\right) _{\alpha }^{\beta }$ to the form $%
c_{\mu }=c\left( \gamma _{\mu }\right) _{\alpha }^{\beta }$ and this implies
that $c=0.$ From the identity
\begin{equation*}
\left[ Q_{\alpha }^{i},\left\{ Q_{\beta }^{j},Q_{\gamma }^{k}\right\} \right]
+\left[ Q_{\beta }^{j},\left\{ Q_{\gamma }^{k},Q_{\alpha }^{i}\right\} %
\right] +\left[ Q_{\gamma }^{k},\left\{ Q_{\alpha }^{i},Q_{\beta
}^{j}\right\} \right] =0,
\end{equation*}
we deduce that $U^{ij}$ and $V^{ij}$ commute with all the generators.
Finally from the identity
\begin{equation*}
\left[ P_{\mu },\left\{ Q_{\alpha }^{i},Q_{\beta }^{j}\right\} \right]
+\left\{ Q_{\alpha }^{i},\left[ Q_{\beta }^{j},P_{\mu }\right] \right\}
-\left\{ Q_{\beta }^{j},\left[ P_{\mu },Q_{\alpha }^{i}\right] \right\} =0,
\end{equation*}
we deduce that $s=0.$ We shall rescale the parameter $r$ to $2.$ Summarizing
the results we have the most general graded algebra which is an extension of
the Poincare symmetery and consistent with unitarity:
\begin{align*}
\left[ Q_{\alpha }^{i},J_{\mu \nu }\right] & =-\frac{i}{2}\left( \gamma
_{\mu \nu }\right) _{\alpha }^{\beta }Q_{\beta }^{i},\quad \left[ Q_{\alpha
}^{i},P_{\mu }\right] =0, \\
\left\{ Q_{\alpha }^{i},Q_{\beta }^{j}\right\} & =-2\left( \gamma ^{\mu
}C\right) _{\alpha \beta }P_{\mu }\delta ^{ij}+C_{\alpha \beta
}U^{ij}+\left( \gamma _{5}C\right) _{\alpha \beta }V^{ij}.
\end{align*}

\subsection{Irreducible representations of supersymmetry}

The Casimir operators are $P^{2}$ and $\widehat{W}^{2}$ where
\begin{equation*}
\widehat{W}_{\mu }=\epsilon _{\mu \nu \rho \sigma }P^{\nu }J^{\rho \sigma }+%
\overline{Q}^{i}\gamma _{\mu }\gamma _{5}Q^{i}.
\end{equation*}
Since $P^{2}$ is a Casimir, all particles belonging to the same multiplet
must have the same mass. Now introduce the fermion number $\left( -1\right)
^{N_{F}}$ which acts on bosonic and fermionic states as follows
\begin{align*}
\left( -1\right) ^{N_{F}}\left| boson\right\rangle & =\left|
boson\right\rangle , \\
\left( -1\right) ^{N_{F}}\left| fermion\right\rangle & =-\left|
fermion\right\rangle ,
\end{align*}
which implies that
\begin{equation*}
\left( -1\right) ^{N_{F}}Q_{\alpha }^{i}\left| .\right\rangle =-Q_{\alpha
}^{i}\left( -1\right) ^{N_{F}}\left| .\right\rangle .
\end{equation*}
For any finite dimensional representation we have
\begin{align*}
Tr\left( \left( -1\right) ^{N_{F}}\left\{ Q_{\alpha }^{i},\overline{Q}%
^{i\beta }\right\} \right) & =Tr\left( \left( -1\right) ^{N_{F}}(Q_{\alpha
}^{i}\overline{Q}^{i\beta }+\overline{Q}^{i\beta }Q_{\alpha }^{i}\right)  \\
& =Tr\left( \left( -1\right) ^{N_{F}}\left( \gamma ^{\mu }\right) _{\alpha
}^{\beta }P_{\mu }\right)  \\
& =Tr\left( -Q_{\alpha }^{i}\left( -1\right) ^{N_{F}}\overline{Q}^{i\beta
}+Q_{\alpha }^{i}\left( -1\right) ^{N_{F}}\overline{Q}^{i\beta }\right) =0.
\end{align*}
From this we deduce that
\begin{equation*}
Tr\left( \left( -1\right) ^{N_{F}}\right) =0,
\end{equation*}
implying that the number of fermionic and bosonic degrees of freedom are
identical.

The energy in supersymmetric theories is non-negative. To see this consider
\begin{align*}
0& \leq \sum\limits_{i}\left( Q_{\alpha }^{i}\left( Q_{\alpha }^{i}\right)
^{\dagger }+\left( Q_{\alpha }^{i}\right) ^{\dagger }Q_{\alpha }^{i}\right)
=\sum\limits_{i}\left( Q_{\alpha }^{i}\left( \overline{Q}^{i}\gamma
_{0}\right) ^{\alpha }+\left( \overline{Q}^{i}\gamma _{0}\right) ^{\alpha
}Q_{\alpha }^{i}\right)  \\
& =Tr\sum\limits_{i}\left\{ Q_{\alpha }^{i},\left( \overline{Q}^{i}\gamma
_{0}\right) ^{\alpha }\right\} =2NTr\left( \gamma ^{\mu }P_{\mu }\gamma
_{0}\right) =8NP_{0}.
\end{align*}
Therefore $E\geq 0.$

We would like to characterize the massless representations of supersymmetry:
$P^{2}=0,$ and we can choose $P_{\mu }=\left( w,0,0,w\right) .$ In this
frame, and after setting the central charges $U^{ij}$ and $V^{ij}$ to zero,
we get
\begin{align*}
\left\{ Q_{\alpha }^{i},Q_{\beta }^{j}\right\} & =2\delta ^{ij}\left( \gamma
^{\mu }C\right) _{\alpha \beta }P_{\mu }=2w\delta ^{ij}\left( \left( \gamma
_{0}+\gamma _{3}\right) C\right)  \\
& =-4w\delta ^{ij}\left(
\begin{array}{cccc}
0 & 0 & 0 & 1 \\
0 & 0 & 0 & 0 \\
0 & 0 & 0 & 0 \\
1 & 0 & 0 & 0
\end{array}
\right) .
\end{align*}
We therefore have one independent relation $\left\{
Q_{1}^{i},Q_{4}^{j}\right\} =-4E\delta ^{ij}.$ From the relation $Q_{\alpha
}^{i}=C_{\alpha \beta }\overline{Q}^{i\beta }$ we have
\begin{equation*}
Q_{3}^{i}=Q_{2}^{i\dagger },\quad Q_{4}^{i}=-Q_{1}^{i\dagger }
\end{equation*}
so we can write $\left\{ Q_{1}^{i},Q_{1}^{j\dagger }\right\} =4w\delta ^{ij}.
$ Normalizing the $Q_{1}^{i}$ we can finally write
\begin{equation*}
\left\{ Q_{1}^{i},Q_{1}^{j\dagger }\right\} =\delta ^{ij}.
\end{equation*}
The subgroup of the Lorentz group leaving the state $\left( w,0,0,w\right) $
invariant is $\Lambda _{03}=0,$ $\Lambda _{10}=-\Lambda _{13},$ $\Lambda
_{20}=-\Lambda _{23}$ implying that the generators are
\begin{equation*}
T_{1}=J_{10}+J_{13},\quad T_{2}=J_{20}+J_{23},\quad J=J_{12}.
\end{equation*}
These obey the commutation relations
\begin{align*}
\left[ J,T_{1}\right] & =T_{2},\quad \left[ J,T_{2}\right] =-T_{1},\quad
\left[ T_{1},T_{2}\right] =0, \\
\left[ J,Q_{1}^{i}\right] & =\frac{1}{2}Q_{1}^{i},\quad \left[ J,\overline{Q}%
_{1}^{i}\right] =-\frac{1}{2}\overline{Q}_{1}^{i},\quad \left\{
Q_{1}^{i},Q_{1}^{j\dagger }\right\} =\delta ^{ij}.
\end{align*}
Since $Q_{1}^{i}$ and $Q_{1}^{i\dagger }$ form a Clifford algebra, they
correspond to a set of $N\;$fermi creation operators $Q_{1}^{i\dagger }$ and
$N$ fermi annihilation operators $Q_{1}^{i}.$ The vacuum state is defined by
$\left| \lambda \right\rangle $%
\begin{align*}
Q_{1}^{i}\left| \lambda \right\rangle & =0,\quad J\left| \lambda
\right\rangle =\lambda \left| \lambda \right\rangle , \\
Q_{1}^{i\dagger }\left| \lambda \right\rangle & =\left| \lambda -\frac{1}{2}%
,i\right\rangle ,\quad Q_{1}^{i\dagger }Q_{1}^{j\dagger }\left| \lambda
\right\rangle =\left| \lambda -1,\left[ ij\right] \right\rangle , \\
Q_{1}^{1\dagger }Q_{1}^{2\dagger }\cdots Q_{1}^{N\dagger }\left| \lambda
\right\rangle & =\left| \lambda -\frac{N}{2}\right\rangle .
\end{align*}
These form $2^{N}$ states falling into two classes. Each class has $2^{N-1}$
states  obtained from $\left| \lambda \right\rangle $ by acting with odd and
even number of $Q_{1}^{i\dagger }$ respectively. The $2^{N}$ states have
helicities ranging from $\lambda _{\max }$ to $\lambda _{\max }-\frac{N}{2}.$
By the CPT theorem physical states must have both helicities $\lambda $ and $%
-\lambda $ \ present which implies that $\lambda _{\max }-\frac{N}{2}%
=-\lambda _{\max }$ and therefore $\lambda _{\max }=\frac{N}{4}.$ If this is
not satisfied one must add the conjugate states where $\lambda _{\max
}^{\prime }=\frac{N}{2}-\lambda _{\max }$ and act on this state with $%
Q_{1}^{i\dagger }.$ A massless non CPT conjugate multiplet has $2^{N+1}$
states. For gauge and matter theories $\lambda _{\max }\leq 1$ which implies
that $N\leq 4.$ We can construct the following table:

\begin{equation*}
\begin{array}{|c|c|c|c|c|c|c|}
\hline
& I & II & III & IV & V & VI\nonumber \\ \hline
\lambda  & N=1 & N=2 & N=1 & N=2 & N=3 & N=4\nonumber \\ \hline
1 &  & 1 &  & 1 & 1 & 1\nonumber \\ \hline
\frac{1}{2} & 1 & 1 & 1+1^{\prime } & 2 & 3+1 & 4\nonumber \\ \hline
0 & 1+1 &  & 2+2^{\prime } & 1+1^{\prime } & \overline{3}+3 & 6\nonumber \\
\hline
-\frac{1}{2} & 1 & 1^{\prime } & 1+1^{\prime } & 2 & 1+\overline{3} &
\overline{4}\nonumber \\ \hline
-1 &  & 1 &  & 1 & 1 & 1\nonumber \\ \hline
\end{array}
\end{equation*}
The multiplets in cases I and II have $\lambda _{\max }=\frac{1}{2}$, while
in all the other cases $\lambda _{\max }=1$. Case I corresponds to an $N=1$
chiral multiplet with $2^{1+1}=4$ states. Case II is $N=1$ vector multiplet
also with $2^{1+1}=4$ states. Case III\ is an $N=2$ vector multiplet with $%
2^{2+1}=8$ states. Case IV is an $N=2$ hypermultiplet with $2^{2+1}=8$
states. Finally case VI give the $N=4$ vector multiplet with $2^{4}=16$
states as it is CPT self conjugate. The $N=3$ in case V is identical to  $N=4
$ because one must add the CPT conjugate states.

Next we consider the situation where $\lambda _{\max }=2$ which implies that
$N\leq 8.$ All these multiplets will contain the graviton and will
correspond to supergravity multiplets:
\begin{equation*}
\begin{array}{|c|c|c|c|c|c|c|c|c|}
\hline
\lambda  & N=1 & N=2 & N=3 & N=4 & N=5 & N=6 & N=7 & N=8\nonumber \\ \hline
2 & 1 & 1 & 1 & 1 & 1 & 1 & 1 & 1\nonumber \\ \hline
\frac{3}{2} & 1 & 2 & 3 & 4 & 5 & 6 & 7+1 & 8\nonumber \\ \hline
1 &  & 1 & \overline{3} & 6 & 10+1 & 15 & 21+7 & 28\nonumber \\ \hline
\frac{1}{2} &  &  & 1 & \overline{4} & \overline{10}+1 & 20+1 & 35+21 & 56%
\nonumber \\ \hline
0 &  &  &  & 1+1 & \overline{5}+5 & \overline{6}+6 & \overline{35}+35 & 70%
\nonumber \\ \hline
-\frac{1}{2} &  &  & 1 & 4 & 1+10 & 1+\overline{20} & \overline{21}+%
\overline{35} & \overline{56}\nonumber \\ \hline
-1 &  & 1 & 3 & 6 & \overline{10}+1 & \overline{15} & \overline{21}+%
\overline{7} & \overline{28}\nonumber \\ \hline
-\frac{3}{2} & 1 & 2 & \overline{3} & \overline{4} & \overline{5} &
\overline{6} & \overline{7}+1 & \overline{8}\nonumber \\ \hline
-2 & 1 & 1 & 1 & 1 & 1 & 1 & 1 & 1\nonumber \\ \hline
\end{array}
\end{equation*}

Notice that $N=7$ supergravity is identical to $N=8$ supergravity, so there
are seven different supergravity theories. Only the field content of the
first three supergravity theories is fixed uniquely. The simplest is $N=1$
supergravity where only two fields are needed, the metric (graviton) and the
gravitino ( Rarita-Schwinger).

\subsection{Supersymmetric field theories}

\subsubsection{\protect\bigskip Lagrangian for chiral multiplet}

The Lagrangian for $N=1$ chiral multiplet $\left( A,B,\psi \right) $ is
given by
\begin{equation*}
L=\frac{1}{2}\left( \partial _{\mu }A\partial ^{\mu }A+\partial _{\mu
}B\partial ^{\mu }B+i\overline{\psi }\gamma ^{\mu }\partial _{\mu }\psi
\right) .
\end{equation*}
The action is invariant under the supersymmetry transformations
\begin{align*}
\delta A& =\overline{\epsilon }\psi ,\quad \delta B=\overline{\epsilon }%
\gamma _{5}\psi , \\
\delta \psi & =\left( -i\partial _{\mu }A+\partial _{\mu }B\gamma
_{5}\right) \gamma ^{\mu }\epsilon ,
\end{align*}
as can be easily verified. The supersymmetry parameter $\epsilon $ is
independent of the coordinates $x^{\mu }$ and supersymmetry is a global
symmetry.

\subsubsection{Lagrangian for vector multiplet}

The vecor multiplet is given by $\left( A^{\mu },\lambda \right) $ where $%
A_{\mu }=A_{\mu }^{a}T^{a}$ and $\lambda =\lambda ^{a}T^{a}$ which are Lie
algebra valued: $\left[ T^{a},T^{b}\right] =if^{abc}T^{c}.$ The Lagrangian
which is valid in dimensions $D=4,6$ and $10$ is given by
\begin{equation*}
L=Tr\left( -\frac{1}{4}F_{\mu \nu }F^{\mu \nu }+\frac{i}{2}\lambda \Gamma
^{\mu }D_{\mu }\lambda \right) ,
\end{equation*}
where $F=dA+A^{2}=\frac{1}{2}F_{\mu \nu }dx^{\mu }\wedge dx^{\nu },$ $F_{\mu
\nu }=F_{\mu \nu }^{a}T^{a}$ and $D_{\mu }\lambda =\partial _{\mu }\lambda
+ig\left[ A_{\mu },\lambda \right] =D_{\mu }\lambda ^{a}T^{a}.$ The action
is invariant under the supersymmetry transformations
\begin{align*}
\delta A_{\mu }& =i\epsilon \Gamma _{\mu }\lambda , \\
\delta \lambda & =\frac{1}{2}F_{\mu \nu }\Gamma ^{\mu \nu }\epsilon .
\end{align*}
To verify supersymmetry one has to use Fierz identities for the fermions
which are valid only in dimensions $4,6$ and $10.$

One can couple  vector multiplets to chiral multiplets and have a
generalization of the standard model. In the supersymmetric standard model
it turned out that all known fields must have partners, e.g., the electron
must have a bosonic partner, the s-electron, while gauge fields $A_{\mu }$
will have as partners the gauginos $\lambda .$ As we have seen, the fermions
in the standard model are chiral. The chirality condition cannot be
satisfied for theories with more than $N=1$ supersymmetry. In other words,
for supersymmetry to be of any relevance, the Lagrangian at low energies
must have $N=1$ supersymmetry. However, in the real world supersymmetry is
not present because no partners for the observed particles with the same
mass are found, so it must be broken. One of the main advantages of
supersymmetry is that it has good quantum behaviour, where divergencies in
the renormalization of the bosonic fields are cancelled by those coming from
the fermions. To preserve good quantum behaviour, the symmetry must be
broken spontaneously. The Higgs mechanism which is essential in the bosonic
case must also be present here. The Goldstone phenomena where the scalar
fields associated with the broken generators are absorbed by the massless
vector fields to become massive must be generalized. In this case the field
associated with the broken supersymmetry generator, the Goldstino, must be
absorbed by a massless vector-spinor field. In the discussion in the last
section we have seen that this is the Rartia-Schwinger field. This implies
that the supergravity multiplet must be coupled to the matter Lagrangian of
the standard model in such a way that supersymmetry is broken spontaneously,
so that the gravitino absorb one fermionic field to become massive.  In
supergravity the gravitational multiplet plays a role in the breaking of
supersymmetry and this can be used to induce the electroweak breaking in the
standard model. This is a novel phenomena as this implies that physics of
the extremely weak gravitational field influences the low-energy sector.

\subsubsection{N=1 supergravity Lagrangian}

The easiest way to construct the supergravity Lagrangian is to generalize
the construction of the Einstein-Hilbert action based on gauging the
Poincare algebra. Here we gauge the supersymmetry algebra instead. Let
\begin{equation*}
D_{\mu }=\partial _{\mu }+\omega _{\mu }^{\,\,ab}J_{ab}+e_{\mu
}^{a}P_{a}+\psi _{\mu }^{\alpha }Q_{\alpha },
\end{equation*}
be the connection associated with the supersymmetry algebra. The curvature
is easily computed to be
\begin{equation*}
\left[ D_{\mu },D_{\nu }\right] =R_{\mu \nu }^{\quad ab}J_{ab}+T_{\mu \nu
}^{a}P_{a}+\psi _{\mu \nu }^{\alpha }Q_{\alpha },
\end{equation*}
where
\begin{align*}
R_{\mu \nu }^{\quad ab}& =\partial _{\mu }\omega _{\nu }^{\,\,ab}-\partial
_{\nu }\omega _{\mu }^{\,\,ab}+\omega _{\mu }^{\,\,ac}\omega _{\nu c}^{\quad
b}-\omega _{\nu }^{\,\,ac}\omega _{\mu c}^{\quad b}, \\
T_{\mu \nu }^{a}& =\partial _{\mu }e_{\nu }^{a}-\partial _{\nu }e_{\mu
}^{a}+\omega _{\mu }^{\,\,ab}e_{\nu b}-\omega _{\nu }^{\,\,ab}e_{\mu b}+%
\overline{\psi }_{\mu }\gamma ^{a}\psi _{\nu }, \\
\psi _{\mu \nu }& =\left( \partial _{\mu }+\frac{1}{4}\omega _{\mu
}^{\,\,ab}\gamma _{ab}\right) \psi _{\nu }-\left( \partial _{\nu }+\frac{1}{4%
}\omega _{\nu }^{\,\,ab}\gamma _{ab}\right) \psi _{\mu },
\end{align*}
and the action of $J_{ab}$ on the spinor $\psi _{\mu }$ is represented by $%
\frac{1}{4}\gamma _{ab}.$ To proceed one first impose the torsion free
constraint
\begin{equation*}
T_{\mu \nu }^{a}=0,
\end{equation*}
which can be solved to express $\omega _{\mu }^{\,\,ab}$ in terms of $e_{\mu
}^{a}$ and $\psi _{\mu }$ to obtain
\begin{equation*}
\omega _{\mu }^{\,\,ab}=\omega _{\mu }^{\,\,ab}(e)+\frac{1}{4}\left(
\overline{\psi }_{\mu }\gamma ^{a}\psi ^{b}-\overline{\psi }_{\mu }\gamma
^{b}\psi ^{a}+\overline{\psi }^{a}\gamma _{\mu }\psi ^{b}\right)
\end{equation*}
where $\omega _{\mu }^{\,\,ab}(e)$ is the same expression we had in the
non-supersymmetric case, and indices are changed from flat to curved by
using the vierbein $e_{\mu }^{a}$ and its inverse $e_{a}^{\mu }.$ The
supersymmetry transformations are given by
\begin{align*}
\delta e_{\mu }^{a}& =\overline{\epsilon }\gamma ^{a}\psi _{\mu }, \\
\delta \omega _{\mu }^{\,\,ab}& =0, \\
\delta \psi _{\mu }& =\left( \partial _{\mu }+\frac{1}{4}\omega _{\mu
}^{\,\,ab}\gamma _{ab}\right) \epsilon .
\end{align*}
The torsion constraint is not preserved under these transformations. We can
preserve the torsion constraint by modifying the $\delta \omega _{\mu
}^{\,\,ab}$ accordingly. The result is
\begin{equation*}
\delta ^{\prime }\omega _{\mu ab}=-\frac{1}{2}e_{a}^{\nu }e_{b}^{\rho
}\left( \overline{\epsilon }\gamma _{\rho }\psi _{\mu \nu }-\overline{%
\epsilon }\gamma _{\mu }\psi _{\nu \rho }+\overline{\epsilon }\gamma _{\nu
}\psi _{\rho \mu }\right) .
\end{equation*}
The Lagrangian invariant under the supersymmetry transformations is given by
\begin{equation*}
e^{-1}L_{SG}=-\frac{1}{4}R+\overline{\psi }_{\mu }\gamma ^{\mu \nu \rho
}\left( \partial _{\nu }+\frac{1}{4}\omega _{\nu }^{\,\,ab}\gamma
_{ab}\right) \psi _{\rho }
\end{equation*}
The proof of invariance is simplified by using that $eR=\frac{1}{2}\epsilon
^{\mu \nu \rho \sigma }\epsilon _{abcd}e_{\mu }^{a}e_{\nu }^{b}R_{\rho
\sigma }^{\quad cd}$ and by writing the varriation with respect to $\omega
_{\mu }^{\,\,ab}$ in the form $\frac{\delta L}{\delta \omega _{\mu }^{\,\,ab}%
}\delta ^{\prime }\omega _{\mu }^{\,\,ab}$. Since $\omega _{\mu }^{\,\,ab}$
appears linearly and quadratically one can easily see that, after
integrating by parts, its varriation $\frac{\delta L}{\delta \omega _{\mu
}^{\,\,ab}}$ vanishes if the torsion vanishes and therefore there is no need
to substitute the very complicated expression for $\delta ^{\prime }\omega
_{\mu }^{\,\,ab}.$ This method is known as the 1.5 formalism and was used
widely in the early formulations of supergravity because it drastically
simplifies the proof of the supersymmetry invariance of the proposed actions.

\section{Higher dimensional theories}

By matching the bosonic and fermionic degrees of freedom we shall determine
what are the possible supersymmetric multiplets in all possible dimensions.
First a massless vector field in $D$ dimensions have $D-2$ degrees of
freedom as the $A_{0}$ component does not propagate because it has no time
derivatives in the action and one of the transverse components of $A_{i}$
can be gauged away. For fermions a Dirac spinor has $2^{\left[ \frac{D}{2}%
\right] }$ components, but in certain cases could be subjected to either the
Weyl \ chirality condition or to the Majorana condition or both. The first
step is to define the Clifford algebra in dimensions $D$. The Dirac gammas
matrices $\Gamma ^{M}$ where $M=0,1,\cdots ,D-1$ will be $2^{\left[ \frac{D}{%
2}\right] }\times 2^{\left[ \frac{D}{2}\right] }$ matricess. In even
dimensions we can define the gamma matrix $\Gamma _{D+1}=\eta \Gamma
^{0}\Gamma ^{1}\cdots \Gamma ^{D-1}$ where $\left( \Gamma _{D+1}\right)
^{2}=1$ so that $\eta ^{2}=\left( -1\right) ^{s-t}=1$ $\left( mod%
8\right) $,  where $s$ is the number of space-like coordinates and $t$ the
number of time like coordinates. For our considerations we will always take $%
s=D-1$ and $t=1.$ Therefore $\left( \Gamma _{D+1}\right) ^{2}=1$ and one can
always impose the Weyl condition $\Gamma _{D+1}\psi _{\pm }=\pm \psi _{\pm }$%
. No Weyl fermions can exist in odd dimensions. The Majorana
condition $\psi =C\overline{\psi }$ requires the existence of
imaginary representations of the gamma matrices in $D$ dimensions
so that $\Gamma ^{M^{\ast }}=-C\Gamma ^{M}C^{-1}.$ This is only
possible in $D=2,3,4$ $\left( mod 8\right) .$ It is also possible
to impose both the Weyl condition and the Majorana condition
simultaneously only in $D=2$ $\left( mod 8\right) .$ Therefore a
spinor $\lambda $ have $2^{2}\times \frac{1}{2}=2$ components in
$4$ dimensions as one imposes the Majorana condition.
(Supersymmetric multiplets are always defined with respect to
Majorana spinors so the Weyl
condition can only be imposed in dimensions $2$ $\left( mod 8\right) $%
. The number of components for $\lambda $ in $5$ dimensions is $4$ while in $%
6$ dimensions it is $2^{3}\times \frac{1}{2}=4$ components. In dimensions $7$
and $8$ it is $8$ components. In dimension $9$ it is $16$ (although in this
case it is possible to modify the Majorana condition to reduce it to $8$
components). In dimension $10$ one can impose simultaneously the Weyl and
Majorana conditions so that the number of independent components of $\lambda
$ is $8.$ We notice that the number of independent components of $A_{M}$ and
$\lambda $ match in $D=3,4,6$ and $10$ and this explains why the super
Yang-Mills action is invariant under supersymmetry transformations only in
those dimensions. In dimensions higher than $10$ the number of independent
fermionic components increase much faster than the number of independent
components for the vectors $A_{M}$ and supersymmetric actions for the pure
vector multiplet do not exist. This is to be expected because when these
higher dimensional theories are analyzed as viewd in four-dimensions, they
will correspond to vector multiplets with supersymmetry higher than $N=4$
which is not possible from our analysis of the supersymmetry representations.

To analyse supergravity theories in higher dimensions we must first
determine the number of independent degrees of freedom for the graviton.
This is essentially a symmetric metric with $\frac{D\left( D+1\right) }{2}$
components where the components $g_{0i\text{ }}$do not propagate because
they do not acquire second time derivatives. From the other components $%
g_{ij}$ the $D$ diffeomorphism parameters could be used to reduce the
components to $\frac{D\left( D-3\right) }{2}=\frac{\left( D-2\right) \left(
D-1\right) }{2}-1$ so essentially the physical degrees are those of a
symmetric traceless metric in $\left( D-2\right) $ dimensions. From the
Rarita-Schwinger equation for a free field, $\Gamma ^{MNP}\partial _{N}\psi
_{P}=0,$ we can easily show by first contracting this equation with $\Gamma
^{M}$ and making the gauge choice $\Gamma ^{M}\psi _{M}=0$ that $\Gamma
^{N}\partial _{N}\psi _{M}=0$ and $\partial ^{M}\psi _{M}=0$. Therefore $%
\psi _{M}$ behaves like a massless vector with $D-2$ spinor components and
the condition $\Gamma ^{N}\partial _{N}\psi _{M}=0$ eliminates one more
spinor component to give finally $\left( D-3\right) 2^{\left[ \frac{D}{2}%
\right] }\times r$ where the reduction factor $r$ is $\frac{1}{2}$ if the
Majorana condition is imposed, $r$ is $\frac{1}{4}$ if the Majorana-Weyl
condition is imposed and $r=1$ if no conditions are imposed. From this
analysis it is easy to see that the supergravity multiplet in $D=4$ is the
metric (or vierbein) $e_{\mu }^{a}$ and the gravitino $\psi _{\mu }$ as we
have seen before. In five dimensions, the metric has $5$ degrees of freedom
while the gravitino has $8.$ A vector field has $3$ degrees of freedom so
the number of fermionic and bosonic degrees match if we take the multiplet $%
\left( e_{\mu }^{a},A_{\mu },\psi _{\mu }\right) .$ We can classify all
possible theories in various dimensions, but it is easier to start with
maximal supergravity which is $N=8.$

\subsection{\protect\bigskip N=1 eleven dimensional supergravity}

A spinor in $11$ dimensions have $2^{\left[ \frac{11}{2}\right] }\times
\frac{1}{2}=16$ degrees of freedom because the Majorana condition could be
imposed. As viewed from $4$ dimensions where a spinor have $2$ degrees of
freedom, this corresponds to a spinor with $SO(8)$ symmetry, and according
to the classification discussed before, this corresponds to a theory with $%
N=8$ supersymmetry in four-dimensions. Therefore $D=11$ is the highest
dimensions where a supergravity theory could be constructed. Also in higher
dimensions the number of fermionic degrees increase much faster than those
of the metric and other bosonic fields. To determine the field content of
the supergravity multiplet in $D=1$ we note that the metric has $44$ degrees
of freedom while the gravitino has $128.$ The difference is $84$ degrees
which must be present in other bosonic configuration. An antisymmetric
tensor of rank $p$ (a $p$-form ) has $\left(
\begin{array}{c}
D-2 \\
p
\end{array}
\right) $ degrees of freedon, which is equal to $84$ when $p=3.$ We deduce
that in the multiplet $\left( e_{M}^{A},A_{MNP},\psi _{M}\right) $ the
bosonic and fermionic degrees of freedom match. This is a necessary but not
sufficient condition to have a supersymmetric multiplet and construct a
supersymmetric action. This is indeed possible and the full supersymmetric
Lagrangian is
\begin{align*}
e^{-1}L& =-\frac{1}{4}R-\frac{i}{2}\overline{\psi }_{M}\Gamma
^{MNP}D_{N}\left( \frac{\omega +\widehat{\omega }}{2}\right) \psi _{P}-\frac{%
1}{48}F_{MNPQ}F^{MNPQ} \\
& +\frac{2e^{-1}}{12^{4}}\epsilon ^{M_{1}\cdots M_{8}PQR}F_{M_{1}\cdots
M_{4}}F_{M_{5}\cdots M_{8}}A_{PQR} \\
& +\frac{1}{192}\left( \overline{\psi }_{M}\Gamma ^{MNPQRS}\psi _{N}+12%
\overline{\psi }^{P}\Gamma ^{QR}\psi ^{S}\right) \left( F_{PQRS}+\widehat{F}%
_{PQRS}\right) ,
\end{align*}
where
\begin{align*}
F_{MNPQ}& =\frac{1}{4}\left( \partial _{M}A_{NPQ}-\partial
_{N}A_{PQM}+\partial _{P}A_{QNM}-\partial _{Q}A_{MNP}\right) , \\
\widehat{\omega }_{M}^{\,\,AB}& =\omega _{M}^{\,\,AB}(e)+\frac{1}{4}\left(
\overline{\psi }_{M}\Gamma ^{A}\psi ^{B}-\overline{\psi }_{M}\Gamma ^{B}\psi
^{A}+\overline{\psi }^{A}\Gamma _{M}\psi ^{B}\right) , \\
\widehat{F}_{PQRS}& =F_{PQRS}-3\overline{\psi }_{\left[ P\right. }\Gamma
_{QR}\psi _{\left. S\right] }.
\end{align*}
The action is invariant under the supersymmetry transformations
\begin{align*}
\delta e_{M}^{A}& =-i\overline{\epsilon }\Gamma ^{A}\psi _{M}, \\
\delta \psi _{M}& =\left( \partial _{M}+\frac{1}{4}\widehat{\omega }%
_{M}^{\,\,AB}\Gamma _{AB}+\frac{i}{144}\left( \Gamma _{M}^{\;PQRS}-8\delta
_{M}^{P}\Gamma ^{QRS}\right) \widehat{F}_{PQRS}\right) \epsilon , \\
\delta A_{MNP}& =\frac{3}{2}\overline{\epsilon }\Gamma _{\left[ MN\right.
}\psi _{\left. P\right] }.
\end{align*}
A theory in higher dimensions can produce more complicated content in lower
dimensions either by compactification or by dimensional reduction. In
dimensional reduction one assumes that the fields are independent of the
coordinates in a certain number of internal directions. The action then
reduces to that of a lower dimensional theory with more fields. As an
exmaple a metric in $D$ dimensions reduces in $d$ dimensions to a metric, $%
\left( D-d\right) $ vectors and $\frac{\left( D-d\right) \left( D-d+1\right)
}{2}$ scalars. A vector in $D$ dimensions reduces in $d$ dimensions to a
vector and $\left( D-d\right) $ scalars.

\subsection{\protect\bigskip N=1 ten-dimensional Supergravity and Super
Yang-Mills}

The super Yang-Mills action can be reduced from $10$ dimensions to $4$. The
vector fields $A_{M}^{a}$ reduce in $4$ dimensions to the vector $A_{\mu
}^{a}$ and the $6$ scalars $A_{i}^{a},$ $B_{i}^{a}=A_{i+3}^{a},$ $i=4,5,6.$
The spinors $\lambda _{\widehat{\alpha }}^{a}$ in $10$ dimensions reduce to
four spinors in $4$ dimensions $\lambda _{\alpha K},$ $K=1,\cdots ,4$. The
reduced action takes the form
\begin{align*}
I& =\int d^{4}xTr\left( -\frac{1}{4}F_{\mu \nu }^{a}F^{\mu \nu a}+\frac{1}{2}%
D_{\mu }A_{i}^{a}D^{\mu }A_{i}^{a}+\frac{1}{2}D_{\mu }B_{i}^{a}D^{\mu
}B_{i}^{a}\right.  \\
& \qquad \qquad +\frac{g}{2}\overline{\lambda }_{K}\left[ \alpha
_{KL}^{i}A_{i}+i\gamma _{5}\beta _{KL}^{i}B_{i},\lambda _{L}\right]  \\
& \qquad \qquad \left. +\frac{g^{2}}{4}\left( \left[ A_{i},A_{j}\right] ^{2}+%
\left[ B_{i},B_{j}\right] ^{2}+2\left[ A_{i},B_{j}\right] ^{2}\right)
\right) ,
\end{align*}
where $\alpha _{KL}^{i}$ and $\beta _{KL}^{i}$ are $6$ real antisymmetric $%
4\times 4$ matrices obeying $SU(2)\times SU(2)$ algebra.

Of particular importance is $N=1$ supergravity in $10$ dimensions. The
supermultiplet consists of the vielbein $e_{M}^{A},$ an antisymmetric field $%
B_{MN}$, a dilaton field $\phi $, a gravitino $\psi _{M}$ and a spinor $%
\lambda .$ The supergravity action in ten dimensions can be obtained from
the eleven dimensional action by dimensional reduction and consistent
truncation. The eleven dimensional vierbein gives in ten dimensions a
vielbein, a vector and a scalar dilaton field. The antisymmetric field $%
A_{MNP}$ gives in ten dimensions antisymmetric tensors of ranks three and
two. The gravitino in eleven dimensions give in ten dimensions two
gravitinos and two spinors. A consistent truncation is to set the vector,
the rank three antisymmetric tensor, one of the gravitinos and one of the
spinors to zero. One can show that one of the two supersymmetries will be
preserved by this truncation leaving a theory with $N=1$ supersymmetry. The
untruncated theory has $N=2$ supersymmetry in ten dimensions. The $N=1$
supergravity Lagrangian is given by
\begin{align*}
e^{-1}L& =-\frac{1}{4}R+\frac{1}{12}e^{-2\phi }H_{MNP}H^{MNP}+\frac{1}{2}%
\partial _{M}\phi \partial ^{M}\phi  \\
& -\frac{i}{2}\overline{\psi }_{M}\Gamma ^{MNP}\left( D_{N}+\widehat{D}%
_{N}\right) \psi _{P}+\frac{i}{2}\overline{\chi }\Gamma ^{M}\left( D_{N}+%
\widehat{D}_{N}\right) \chi  \\
& +\frac{1}{2\sqrt{2}}\overline{\psi }_{M}\Gamma ^{N}\Gamma ^{M}\chi \left(
\partial _{N}\phi +\widehat{D}_{N}\phi \right)  \\
& +\frac{i}{48}\left( \overline{\psi }_{M}\Gamma ^{MNPQR}\psi _{R}+6%
\overline{\psi }^{N}\Gamma ^{P}\psi ^{Q}-i\sqrt{2}\overline{\psi }_{M}\Gamma
^{NPQ}\Gamma ^{M}\chi \right) \left( H_{NPQ}+\widehat{H}_{NPQ}\right) ,
\end{align*}
where $H_{MNP}=3\partial _{\left[ M\right. }B_{\left. NP\right] }$. This
action is invariant under the supersymmetry transformations
\begin{align*}
\delta e_{M}^{A}& =-i\overline{\epsilon }\Gamma ^{A}\psi _{M}, \\
\delta \phi & =-\frac{1}{\sqrt{2}}\overline{\epsilon }\chi , \\
\delta B_{MN}& =e^{\phi }\left( i\epsilon \Gamma _{\left[ M\right. }\psi
_{\left. N\right] }\right) , \\
\delta \psi _{M}& =\widehat{D}_{M}\epsilon -\frac{\sqrt{2}}{48}e^{-\phi
}\left( \Gamma _{M}^{\;NPQ}-9\delta _{M}^{N}\Gamma ^{PQ}\right) \widehat{H}%
_{NPQ}, \\
\delta \chi & =\frac{i}{\sqrt{2}}\Gamma ^{M}\epsilon \widehat{D}_{M}\phi +%
\frac{i}{12\sqrt{2}}e^{-\phi }\Gamma ^{NPQ}\widehat{F}_{NPQ}.
\end{align*}
The main advantage of the $N=1$ supergravity Lagrangian in ten dimensions is
that it can be coupled to the ten-dimensional super Yang-Mills theory with
an arbitrary gauge group. The main modification to the pure supergravity
Lagrangian is that the field $H_{MNP}$ is replaced with
\begin{align*}
\widehat{H}_{MNP}& =H_{MNp}-\omega _{MNP}, \\
\omega _{MNP}& =Tr\left( A_{\left[ M\right. }\partial _{N}A_{\left. P\right]
}+\frac{g}{3}A_{\left[ M\right. }A_{N}A_{\left. P\right] }\right) ,
\end{align*}
where $\omega _{MNP}$ is the Chern-Simons form over the gauge Lie algebra.
When higher derivative terms are allowed in the supergravity Lagrangian then
the modifications to $H_{MNP}$ will also include the Chern-Simons
gravitational term. The field $\widehat{H}_{MNP}$ satisfies the equation
\begin{equation*}
d\widehat{H}=tr\left( -F\wedge F+R\wedge R\right) ,
\end{equation*}
where $\widehat{H}$ is a three-form, $F$ is the gauge field strength
two-form and $R$ is the curvature tensor two-form.

\section{Further directions}

The topics considered so far in these lectures would enable us to study some
very interesting questions in physics using results from mathematics. For
lack of space, these directions will only be described briefly. The
interested reader can pursue these problems in the literature.

\subsection{\protect\bigskip Index of Dirac operators}

Consider Dirac equation for a massless spinor in ten-dimensions
\begin{align*}
iD_{10}\psi & =0=i\Gamma ^{M}D_{M}\psi =0, \\
& =i\left( D_{4}+D_{K}\right) \psi ,
\end{align*}
where $K$ is the internal $6$ dimensional manifold. Let $\Gamma ^{\left(
4\right) }=i\Gamma _{0}\Gamma _{1}\Gamma _{2}\Gamma _{3}$ and $\Gamma
^{\left( K\right) }=-i\Gamma _{4}\Gamma _{5}\cdots \Gamma _{9}$ and $\Gamma
^{\left( 10\right) }=\Gamma _{0}\Gamma _{1}\cdots \Gamma _{9}.$ This implies
that $\left( \Gamma ^{\left( 4\right) }\right) ^{2}=\left( \Gamma ^{\left(
K\right) }\right) ^{2}=\left( \Gamma ^{\left( 10\right) }\right) ^{2}=1$ and
$\Gamma ^{\left( 10\right) }=\Gamma ^{\left( 4\right) }\Gamma ^{\left(
K\right) }.$ The Weyl condition on fermions in ten-dimensions $\Gamma
^{\left( 10\right) }\psi =\psi $ is equivalent to $\Gamma ^{\left( 4\right)
}\psi =\Gamma ^{\left( K\right) }\psi .$ Defining $\widehat{D}_{4}=\Gamma
^{\left( 4\right) }D_{4}$ and $\widehat{D}_{K}=\Gamma ^{\left( 4\right)
}D_{K}$ it is easy to see that one can diagonalize simultaneously $\widehat{D%
}_{4}$ and $\widehat{D}_{K}$. Let $H=\left( iD_{K}\right) ^{2}$ and let $%
\psi $ be an eigenstate with energy $E,$ $H\psi =E\psi .$ As $H$ commutes
with $\Gamma ^{\left( K\right) }$ we have
\begin{equation*}
H\left( iD_{K}\psi \right) =E\left( iD_{K}\psi \right) ,
\end{equation*}
so that $\psi $ and $D_{K}\psi $ are degenerate in energy. On the other hand
\begin{equation*}
D_{K}\left( \Gamma ^{\left( K\right) }\psi \right) =-\Gamma ^{\left(
K\right) }D_{K}\psi ,
\end{equation*}
so that $\psi $ and $D_{K}\psi $ have opposite chiralities with respect to $%
\Gamma ^{\left( K\right) }$ and are therefore linearly independent, unless $%
D_{K}\psi =0.$ For every state with $\Gamma ^{\left( K\right) }=1$ there is
a state with $\Gamma ^{\left( K\right) }=-1$ except for zero eigenvalues
which do not have to be paired. The index of the Dirac operator is then
defined by
\begin{equation*}
index\left( iD_{K}\right) =n_{+}-n_{-}
\end{equation*}
where $n_{+}$ and $n_{-}$ are, respectively, the number of zero eigenvalues
with eigenvalues $\Gamma ^{\left( K\right) }=1$ and $\Gamma ^{\left(
K\right) }=-1.$ In the absence of gauge fields the Dirac operator satisfies
the properties
\begin{equation*}
\left( D_{K}\psi \right) ^{\ast }=D_{K}\psi ^{\ast },\qquad \left( \Gamma
^{\left( K\right) }D_{K}\psi \right) ^{\ast }=-\Gamma ^{\left( K\right)
}D_{K}\psi ^{\ast }.
\end{equation*}
This is so because in six-dimensions the gamma matrices are real and $\Gamma
^{\left( K\right) }$ is purely imaginary , and therefore complex conjugation
exchanges $n_{+}$ and $n_{-}$, so that  $n_{+}=$ $n_{-}$ and the index is
zero. In presence of gauge fields, and for spinors in some representation $Q$
of the gauge group we have
\begin{equation*}
index_{Q}\left( iD_{K}\right) =-index_{Q^{\ast }}\left( iD_{K}\right)
\end{equation*}
because complex conjugation exchanges the representation $Q$ with its
complex conjugate $Q^{\ast },$ and this implies that the index is zero if
the representation $Q$ is real or pseudoreal. If $Q$ is complex then the
index is not necessarily zero. In terms of the gauge fields and the
spin-connection of the underlying manifold the index of the Dirac operator
can be computed to be
\begin{equation*}
index_{Q}\left( iD_{K}\right) =\frac{1}{48\left( 2\pi \right) ^{3}}\int
\left( tr_{Q}\left( F\wedge F\wedge F\right) -\frac{1}{8}tr_{Q}\left(
F\right) \wedge tr\left( R\wedge R\right) \right) .
\end{equation*}
One can apply this property to explain physical phenomena. We have seen that
in the standard model of particle physics, there are three families of
particles. In higher dimensional theories it is assumed that particles are
in some large representation of a gauge group which is spontaneously broken
and the four dimensional theory is obtained by compactification from the
higher dimensional one. From the higher dimensional point of view all these
particles are massless as they acquire their small masses (compared with the
high energy scale of compactification) at a later stage. The number of
massless families is then linked to the index of the Dirac operator on the
internal manifold. Therefore we can write
\begin{equation*}
index_{Q}\left( iD_{K}\right) =\frac{1}{2}\chi \left( K\right)
\end{equation*}
where $\chi \left( K\right) $ is the Euler number of the internal manifold.
To build a realistic model with three families one must choose the Euler
number of the internal manifold to be $6.$

\subsection{Holonomy and N=1 supersymmetry}

At low-energies it is desirable to have a field theory with $N=1$
supersymmetry in four-dimensions. To have some unbroken supersymmetry, let $Q
$ be the supersymmetry generator annihilating the vacuum $\left|
0\right\rangle $. This implies that
\begin{equation*}
\left\langle 0\right| \left\{ Q,A\right\} \left| 0\right\rangle =0,
\end{equation*}
for all operators $A.$ If $A$ is bosonic then $\left\{ Q,A\right\} $ is
fermionic, but $\left\langle 0\right| $ fermionic $\left| 0\right\rangle =0$
as a fermionic state has no expectation value. If $A$ is fermionic then $%
\left\{ Q,A\right\} =\delta A$ and the above relation implies that $\delta
A=0.$ In supergravity theories we have seen that the supersymmetry
transformations of the gravitino and the other fermions are proportional to
the other bosonic fields and the supersymmetry parmeter. For example,
assuming that bosonic fields, except for the metric, are set to zero, the
gravitino transformation is given by
\begin{equation*}
\delta \psi _{M}=D_{M}\epsilon =0.
\end{equation*}
But this is a Killing spinor equation and $\epsilon $ is a covariantly
constant spinor. This can be satisfied provided Killing spinors exist on the
manifold. This also imply consistency conditions
\begin{equation*}
\left[ D_{M},D_{N}\right] \epsilon =0,
\end{equation*}
or equivalently $R_{MNPQ}\Gamma ^{PQ}\epsilon =0.$ Multiplying by $\Gamma
^{N}$ and using the property
\begin{equation*}
\Gamma ^{N}\Gamma ^{PQ}=\Gamma ^{NPQ}+g^{NP}\Gamma ^{Q}-g^{NQ}\Gamma ^{P},
\end{equation*}
where $\Gamma ^{NPQ}$ is completely antisymmetric in the three indices, and
using the Riemann identity $R_{M\left[ NPQ\right] }=o,$ we deduce that
\begin{equation*}
R_{MQ}\Gamma ^{Q}\epsilon =0.
\end{equation*}
Let us impose the constraint that the vacuum state of the ten-dimensional
space is of the form $M^{4}\times K$ where $M^{4}$ is a maximally symmetric
four-dimensional space and $K$ is a compact six-dimensional manifold. In
this case the consistency condition on the Riemann tensor does not admit
deSitter or anti deSitter spaces as solutions because in this case
\begin{equation*}
R_{\mu \nu \rho \sigma }=\frac{r}{12}\left( g_{\mu \rho }g\nu \sigma -g_{\mu
\sigma }g_{\nu \rho }\right)
\end{equation*}
which togother with the consistency condition implies that $r=0.$

Upon parallel transport around a contractible closed curve $\gamma $, a
field $f$ is transformed into $Uf$ where
\begin{equation*}
U=P\exp _{\gamma }\int \omega dx
\end{equation*}
where $\omega $ is the spin-connection of the $n$-dimensional internal
manifold $K.$ The $SO(n)$ matrices $U$ obtained this way always form a group
$H$ called the holonomy group. We can then ask : Under what conditions the
manifold $K$ admits a spinor field $\epsilon $ obeying $D_{i}\epsilon =0$ ?
A covariantly constant spinor $\epsilon $ keeps its original value under
parallel transport: $U\epsilon =\epsilon .$ As an example, let us consider
compactifying ten-dimensional supergrravity to four dimensions. In this case
the internal manifold $K$ is $6$ dimensional. The subgroup of the $SO(6)$ is
isomorphic to $SU(4)$ and the left-handed spinor obeying the above condition
will be in the fundamental $4$-representation. By an $SU(4)$ rotation we can
always transform $\epsilon $ to the form
\begin{equation*}
\epsilon =\left(
\begin{array}{c}
0 \\
0 \\
0 \\
\epsilon _{0}
\end{array}
\right)
\end{equation*}
and this is left invariant by the $SU(3)$ subgroup. We can write the
following decomposition
\begin{eqnarray*}
SO(1,9) &\rightarrow &SO(1,3)\times SO(6)\sim SO(1,3)\times SU(4), \\
16 &=&\left( 2,4\right) \oplus (2^{\prime },\overline{4})
\end{eqnarray*}
where we have decomposed a $16$ dimensional spinor of $SO(1,9)$ (the Lorentz
group in ten dimensions) in terms of the four-dimensional spinors of $SO(1,3)
$ and the $4$ and $\overline{4}$ spinor representations of $SO(6).$ From
these considerations we conclude that by requiring the manifold $K$ to have $%
SU(3)$ holonomy, the compactified theory will have $N=1$ supersymmetry in
four-dimensions. Therefore the question to ask now is: What sort of manifold
$K$ admits a metric with $SU(3)$ holonomy?

If $\epsilon $ is a covariantly constant spinor, then so would be the K\"{a}%
hlerian form $k_{ij}=\overline{\epsilon }\Gamma _{ij}\epsilon $, the complex
structure $J_{j}^{i}=g^{il}k_{lj}$ and the volume form $\omega
_{ijk}=\epsilon ^{T}\Gamma _{ijk}\epsilon .$ In six-dimensions, if the
holonomy group is not $SO(6)$ but $U(3)$ then at any point in $K$ tangent
vectors which transform as vectors of $SO(6)$ decompose as $3+\overline{3}$
under $U(3).$ There is a unique matrix $J_{j}^{i}$ acting on tangent vectors
in the $3$ and $\overline{3}$ representations. $J$ defines an almost complex
structure and is invariant under the holonomy group. Therefore it is
covariantly constant and its derivatives vanish. A manifold of $U(N)$
holonomy is complex and is also K\"{a}hler. If we ask whether it is possible
to find a K\"{a}hler manifold with $SU(N)$ holonomy instead of $U(N)$
holonomy, then the answer lies in the Calabi-Yau conjecture. This states
that a K\"{a}hler manifold of vanishing first Chern class admits a K\"{a}%
hler metric of $SU(N)$ holonomy and this metric is unique.

\subsection{Solutions of Einstein equations}

There are advantages to deal with a supersymmetric system when solving
Einstein equations resulting from a coupled gravitational system. This is
done by embedding the bosonic system in a supergravity theory, whenever this
is possible. One first starts with an ansatz for the metric dictated by the
symmetry of the problem. The supersymmetry transformations depend on first
order derivatives of the fields. Setting the supersymmetric transformations
to zero and requiring that the solution preserve all or part of the
supersymmetry, results in a system of first order differential equations, in
contrast to the second order differential equations which one have in the
original bosonic system. This method has been very successful in finding new
solutions for Einstein-coupled systems. To illustrate the power of this
method we consider the follwing model.

Consider the coupling of gravity to a dilaton field $\phi $ and an $SU(2)$
Yang-Mills field $A_{\mu }^{a}.$ described by the action
\begin{equation*}
S=\int \left( -\frac{1}{4}\,R+\frac{1}{2}\,\partial _{\mu }\phi \,\partial
^{\mu }\phi -\frac{1}{4}\,e^{2\phi }\,F_{\mu \nu }^{a}F^{a\mu \nu }+\frac{1}{%
8}\,e^{-2\phi }\right) \sqrt{-\mathbf{g}}\,d^{4}x,
\end{equation*}
where $F_{\mu \nu }^{a}=\partial _{\mu }A_{\nu }^{a}-\partial _{\nu }A_{\mu
}^{a}+\varepsilon _{abc}A_{\mu }^{b}A_{\nu }^{c}$, and the dilaton potential
can be viewed as an effective negative, position-dependent cosmological term
$\Lambda (\phi )=-\frac{1}{4}e^{-2\phi }$.

We shall consider static, spherically symmetric, purely magnetic
configurations of the bosonic fields, and for this we parameterize the
fields as follows:
\begin{equation*}
ds^{2}=N\sigma ^{2}dt^{2}-\frac{dr^{2}}{N}-r^{2}(d\theta ^{2}+\sin
^{2}\theta \,d\varphi ^{2}),
\end{equation*}
\begin{equation*}
\alpha ^{a}A_{\mu }^{a}dx^{\mu }=w\ (-\alpha ^{2}\,d\theta +\alpha
^{1}\,\sin \theta \,d\varphi )+\alpha ^{3}\,\cos \theta \,d\varphi ,
\end{equation*}
where $N$, $\sigma $, $w$, as well as the dilaton $\phi $, are functions of
the radial coordinate $r$. The field equations, following from the action,
read
\begin{equation*}
(rN)^{\prime }+r^{2}N\phi ^{\prime \,\,2}+U+r^{2}\Lambda (\phi )=1,
\end{equation*}
\begin{equation*}
\left( \sigma Nr^{2}\phi ^{\prime }\right) ^{\prime }=\sigma
\,(U-r^{2}\Lambda (\phi )),
\end{equation*}
\begin{equation*}
r^{2}\left( N\sigma e^{2\phi }\,w^{\prime }\right) ^{\prime }=\sigma
e^{2\phi }\,w(w^{2}-1),
\end{equation*}
\begin{equation*}
\sigma ^{\prime }=\sigma \,(r\phi ^{\prime \,\,2}+2e^{2\phi }\,w^{\prime
\,2}/r),
\end{equation*}
where $U=2e^{2\phi }\left( Nw^{\prime \,2}+(w^{2}-1)^{2}/2r^{2}\right) $.
This is a system of second order differential equations which could not be
solved. We now adopt the strategy of embedding this action into a
supersymmetric action and use proporties of supersymmetry to find a solution.

The action of the N=4 gauged SU(2)$\times $SU(2) supergravity includes a
vierbein $e_{\mu }^{\alpha }$, four Majorana spin-3/2 fields $\psi _{\mu
}\equiv \psi _{\mu }^{\mathrm{I}}$ $(\mathrm{{I}=1,\ldots 4)}$, vector and
pseudovector non-Abelian gauge fields $A_{\mu }^{(1)\,a}$ and $A_{\mu
}^{(2)\,a}$ with independent gauge coupling constants $g_{1}$ and $g_{2}$,
respectively, four Majorana spin-1/2 fields $\chi \equiv \chi ^{\mathrm{I}}$%
, the axion $\mathbf{a}$ and the dilaton $\phi $ . The bosonic part of the
action reads
\begin{equation*}
S=\int \left( -\frac{1}{4}\,R+\frac{1}{2}\,\partial _{\mu }\phi \,\partial
^{\mu }\phi +\frac{1}{2}\,e^{-4\phi }\partial _{\mu }\mathbf{a}\,\partial
^{\mu }\mathbf{a}-\frac{1}{4}\,e^{2\phi }\,\sum_{s=1}^{2}{F}_{\mu \nu }^{({s}%
)\,a}F^{({s})\,a\mu \nu }\right.
\end{equation*}
\begin{equation*}
\left. -\frac{1}{2}\,\mathbf{a}\,\sum_{s=1}^{2}{F}_{\mu \nu }^{({s})\,a}\ast
\!F^{({s})\,a\mu \nu }+\frac{g^{2}}{8}\,e^{-2\phi }\right) \sqrt{-\mathbf{g}}%
\,d^{4}x.
\end{equation*}
Here $g^{2}=g_{1}^{2}+g_{2}^{2}$, the gauge field tensor $F_{\mu \nu }^{({s}%
)\,a}=\partial _{\mu }A_{\nu }^{({s})\,a}-\partial _{\nu }A_{\mu }^{({s}%
)\,a}+g_{C}\,\varepsilon _{abc}\,A_{\mu }^{({s})\,b}A_{\nu }^{({s})\,c}$
(there is no summation over ${s}=1,2$), and $\ast \!F_{\mu \nu }^{({s})\,a}$
is the dual tensor. The dilaton potential can be viewed as an effective
negative, position-dependent cosmological term $\Lambda (\phi )=-\frac{1}{8}%
\,g^{2}\,e^{-2\phi }$. The ungauged version of the theory corresponds to the
case where $g_{1}=g_{2}=0$. We consider the truncated theory specified by
the conditions $g_{2}=A_{\nu }^{a\left( 2\right) }=0$. In addition, we
require the vector field $A_{\mu }^{a}$ to be purely magnetic, which allows
us to set the axion to zero.

For a purely bosonic configuration, the supersymmetry transformation laws
are
\begin{equation*}
\delta \bar{\chi}=-\frac{i}{\sqrt{2}}\,\bar{\epsilon}\,\gamma ^{\mu
}\partial _{\mu }\phi -\frac{1}{2}e^{\phi }\,\bar{\epsilon}\,\alpha
^{a}F_{\mu \nu }^{a}\,\sigma ^{\mu \nu }+\frac{1}{4}\,e^{-\phi }\,\bar{%
\epsilon},
\end{equation*}
\begin{equation*}
\delta \bar{\psi}_{\rho }=\bar{\epsilon}\left( \overleftarrow{\partial }%
_{\rho }-\frac{1}{2}\,\omega _{\rho mn}\,\sigma ^{mn}+\frac{1}{2}\,\alpha
^{a}A_{\rho }^{a}\right) -\frac{1}{2\sqrt{2}}\,e^{\phi }\,\bar{\epsilon}%
\,\alpha ^{a}F_{\mu \nu }^{a}\,\gamma _{\rho }\,\sigma ^{\mu \nu }+\frac{i}{4%
\sqrt{2}}\,e^{-\phi }\,\bar{\epsilon}\,\gamma _{\rho },
\end{equation*}
the variations of the bosonic fields being zero. In these formulas, $%
\epsilon \equiv \epsilon ^{\mathrm{I}}$ are four Majorana spinor
supersymmetry parameters, $\alpha ^{a}\equiv \alpha _{\mathrm{IJ}}^{a}$ are
the SU(2) gauge group generators, and $\omega _{\rho mn}$ is the tetrad
connection.

The field configuration is supersymmetric, provided that there are
non-trivial supersymmetry Killing spinors $\epsilon $ for which the
variations of the fermion fields vanish. Specifying to the above
configuration and putting $\delta \bar{\chi}=\delta \bar{\psi}_{\mu }=0$,
the supersymmetry constraints become a system of equations for the four
spinors $\epsilon ^{\mathrm{I}}$. The procedure which solves these equations
is rather involved. The consistency of the algebraic constraints requires
that the determinants of the corresponding coefficient matrices vanish and
that the matrices commute with each other. These consistency conditions can
be expressed by the following relations for the background:
\begin{equation*}
N\sigma ^{2}=e^{2(\phi -\phi _{0})},
\end{equation*}
\begin{equation*}
N=\frac{1+w^{2}}{2}+e^{2\phi }\,\frac{(w^{2}-1)^{2}}{2r^{2}}+\frac{r^{2}}{8}%
e^{-2\phi },
\end{equation*}
\begin{equation*}
r\phi ^{\prime }=\frac{r^{2}}{8N}e^{-2\phi }\left( 1-4e^{4\phi }\,\frac{%
(w^{2}-1)^{2}}{r^{4}}\right) ,
\end{equation*}
\begin{equation*}
rw^{\prime }=-2w\frac{r^{2}}{8N}e^{-2\phi }\left( 1+2e^{2\phi }\frac{w^{2}-1%
}{r^{2}}\right) ,
\end{equation*}
with constant $\phi _{0}$. Under these conditions, the solution of the
algebraic constraints yields $\epsilon $ in terms of only two independent
functions of $r$. The remaining differential constraint then uniquely
specify these two functions up to two integration constants, which finally
corresponds to two unbroken supersymmetries. Introducing the new variables $%
x=w^{2}$ and $R^{2}=\frac{1}{2}r^{2}e^{-2\phi }$, the above equations become
equivalent to one first order differential equation
\begin{equation*}
2xR\,(R^{2}+x-1)\frac{dR}{dx}+(x+1)\,R^{2}+(x-1)^{2}=0.
\end{equation*}
If $R(x)$ is known, the radial dependence of the functions, $x(r)$ and $R(r)$%
, can be determined. Define the following substitution:
\begin{equation*}
x=\rho ^{2}\,e^{\xi (\rho )},\ \ \ \ \ \ \ R^{2}=-\rho \frac{d\xi (\rho )}{%
d\rho }-\rho ^{2}\,e^{\xi (\rho )}-1,
\end{equation*}
where $\xi (\rho )$ is obtained from
\begin{equation*}
\frac{d^{2}\xi (\rho )}{d\rho ^{2}}=2\,e^{\xi (\rho )}.
\end{equation*}
The most general (up to reparametrizations) solution of this equation which
ensures that $R^{2}>0$ is $\xi (\rho )=-2\ln \sinh (\rho -\rho _{0})$. This
gives us the general solution. The metric is non-singular at the origin if
only $\rho _{0}=0$, in which case
\begin{equation*}
R^{2}(\rho )=2\rho \coth \rho -\frac{\rho ^{2}}{\sinh ^{2}\rho }-1,
\end{equation*}
one has $R^{2}(\rho )=\rho ^{2}+O(\rho ^{4})$ as $\rho \rightarrow 0$, and $%
R^{2}(\rho )=2\rho +O(1)$ as $\rho \rightarrow \infty $. The last step is to
obtain $r(s)$, which finally gives us a family of completely regular
solutions of the Bogomolny equations:
\begin{equation*}
d{s}^{2}=a^{2}\,\frac{\sinh \rho }{R(\rho )}\left\{ dt^{2}-d\rho
^{2}-R^{2}(\rho )(d\vartheta ^{2}+\sin ^{2}\vartheta d\varphi ^{2})\right\} ,
\end{equation*}
\begin{equation*}
w=\pm \frac{\rho }{\sinh \rho },\ \ \ \ e^{2\phi }=a^{2}\,\frac{\sinh \rho }{%
2\,R(\rho )},
\end{equation*}

\noindent where $0\leq \rho <\infty $, and we have chosen $2\phi _{0}=-\ln 2$%
. The geometry described by the above line element is everywhere regular,
the coordinates covering the whole space whose topology is R$^{4}$. The
geometry becomes flat at the origin, but asymptotically it is not flat, even
though the cosmological term $\Lambda (\phi )$ vanishes at infinity.  In the
asymptotic region all curvature invariants tend to zero, however, not fast
enough. The Schwarzschild metric functions for $r\rightarrow \infty $ are $%
N\propto \ln r$ and $N\sigma ^{2}\propto r^{2}/4\ln r$, the non-vanishing
Weyl tensor invariant being $\Psi _{2}\propto -1/6r^{2}$.

\end{document}